\documentclass[12pt]{article}

\usepackage{amsthm, enumitem, thmtools}
\usepackage{amssymb}
\usepackage{amsmath,amsfonts}
\usepackage{algorithmic}
\usepackage{hyperref}
\usepackage{array}
\usepackage{subcaption}
\usepackage{tabularx}
\usepackage{multirow}
\usepackage{textcomp}
\usepackage{stfloats}
\usepackage{url}
\usepackage{verbatim}
\usepackage{graphicx, color}
\usepackage{xcolor}
\usepackage[linesnumbered,ruled]{algorithm2e}
\usepackage{authblk}

\newcommand{\up}{\Omega}
\newcommand{\e}{{\mathrm \epsilon}}
\newcommand{\dd}{{\mathrm d}}

\definecolor{amethyst}{rgb}{0.6, 0.4, 0.8}
\definecolor{applegreen}{rgb}{0.55, 0.71, 0.0}
\definecolor{aqua}{rgb}{0.0, 1.0, 1.0}
\definecolor{asparagus}{rgb}{0.53, 0.66, 0.42}
\definecolor{armygreen}{rgb}{0.29, 0.33, 0.13}
\definecolor{custombrown}{rgb}{0.43, 0.21, 0.1}
\definecolor{brightpink}{rgb}{1.0, 0.0, 0.5}
\definecolor{brightube}{rgb}{0.82, 0.62, 0.91}
\definecolor{byzantine}{rgb}{0.74, 0.2, 0.64}

\usepackage{wrapfig}
\definecolor{fig}{RGB}{76,78,92}
\usepackage[labelfont=bf,font={color=fig,small}]{caption}

\usepackage{tikz}
\usetikzlibrary{intersections}
\usetikzlibrary{arrows,snakes,shapes}
\usepackage{pgfplots}
\usepgfplotslibrary{fillbetween}
                
\newcommand{\proton}[1]{%
    \shade[ball color=red] (#1) circle (.25);\draw (#1) node{$+$};
}
\newcommand{\neutron}[1]{%
    \shade[ball color=green] (#1) circle (.25);
}
\newcommand{\electron}[3]{%
    \draw[rotate = #3](0,0) ellipse (#1 and #2)[color=gray];
    \shade[ball color=yellow] (0,#2)[rotate=#3] circle (.1);
}
\newcommand{\nucleus}{%
    \neutron{0.1,0.3}
    \proton{0,0}
    \neutron{0.3,0.2}
    \proton{-0.2,0.1}
    \neutron{-0.1,0.3}
    \proton{0.2,-0.15}
    \neutron{-0.05,-0.12}
    \proton{0.17,0.21}
}

\newcommand{\inelastic}[2]{
  \proton{#1,#2};
  \draw[->,thick,blue](#1+0.5,#2)--(4,-3.7);%
  \draw[->,thick,blue](0,-3)--(-0.3,-4);%
  \shade[ball color=yellow] (-0.3,-4) circle (.1); 
}

\newcommand{\elastic}[2]{
  \proton{#1,#2};
  \draw[->,thick,orange,bend right=90](#1+0.5,#2) to  [out=-30, in=-150] (4,3.);
}

\newcommand{\protoncollision}[3]{
  \proton{#1,#2};
  \draw[->,thick,red](#1+0.5,#2)--(-0.5,0);%
  \draw[->,thick,red](#1+0.5,#2)--(-0.5,0);%
  \draw[->,thick,red](0.5,0)--(3.7,-1.8);%
  \proton{4,-2};  
}
\usepackage{xspace}

\usepackage{tikz, pgfplots}
\pgfplotsset{compat=newest}

\usepackage{natbib}

\title{SDE-based Monte Carlo dose calculation for proton therapy validated against Geant4}

\author[1,*]{Christopher B.C. Dean}
\author[1,2,*]{Maria L. Pérez-Lara}
\author[1]{Emma Horton}
\author[3]{Matthew Southerby}
\author[1,4]{Jere Koskela}
\author[1,**]{Andreas E. Kyprianou}
\affil[1]{Department of Statistics, University of Warwick, Coventry CV4 7AL, United Kingdom.}
\affil[2]{Department of Medical Physics \& Biomedical Engineering, University College London, Gower Street, London WC1E 6BT, United Kingdom.}
\affil[3]{Department of Radiotherapy Physics, University College London Hospitals NHS Foundation Trust, 250 Euston Road, London NW1 2BU, United Kingdom.}
\affil[4]{School of Mathematics, Statistics and Physics, Newcastle University, Newcastle upon Tyne NE1 7RU, United Kingdom.}
\affil[*]{These authors contributed equally to this work.}
\affil[**]{Corresponding author. E-mail: andreas.kyprianou@warwick.ac.uk.}

\makeatletter
\renewcommand\AB@affilsepx{, \protect\footnotemark}
\makeatother

\date{}

\begin{document}

\maketitle
\begin{abstract}

\textit{Objective: }To systematically assess the accuracy and computational performance of a newly proposed stochastic differential equation (SDE)-based model for proton beam dose calculation by benchmarking it against Geant4 in a set of simplified but increasingly challenging phantom geometries.

\textit{Approach: }Building on previous work in \cite{SDE}, where energy deposition from a proton beam was modelled using an SDE framework, we implemented the model using standard approximations to interaction cross sections and mean excitation energies, enabling straightforward adaptation to new materials and configurations. The model was benchmarked against Geant4 in homogeneous, longitudinally heterogeneous and laterally heterogeneous phantoms, for assessment of depth-dose behaviour, lateral transport and impact of material heterogeneities.

\textit{Main results: }Across all phantom configurations and beam energies, the SDE model reproduced the main depth–dose characteristics predicted by Geant4, with proton range agreement within 0.2 mm for 100 MeV beams and within 0.6 mm for 150 MeV beams. Voxel-wise comparisons yielded gamma pass rates exceeding 95\% for all cases under strict 2\%/0.5 mm criteria with a 1\% dose threshold. Differences between the two approaches were spatially localised and primarily associated with regions of steep dose gradients or material heterogeneities, while overall lateral beam dispersion was well reproduced. In terms of computational performance, the SDE model achieved speed-up factors of approximately 2.5--3 relative to single-threaded Geant4, consistently across different Geant4 physics lists.

\textit{Significance: }These results demonstrate that the SDE-based approach can reproduce key dosimetric features predicted by high-fidelity Monte Carlo simulations with good accuracy while already offering a moderate reduction in computational cost. Owing to its formulation, the method is naturally amenable to parallel and GPU-accelerated implementations, suggesting potential for substantial further speed improvements. This makes the approach a promising candidate for fast dose calculations in proton therapy.
\vskip 11pt


 \end{abstract}

\section{Introduction}

Accurate dose calculations are critical for proton therapy treatment planning to ensure precise treatment delivery while minimising uncertainties in patient outcomes. Thus, high-fidelity models are needed to capture the inherently stochastic nature of proton interactions in tissue, including energy loss, scattering, and nuclear interactions. Currently, pencil beam algorithms \citep{hong1996pencil, schaffner1999dose} remain widely used in clinical workflows due to their computational efficiency, particularly for treatment planning and optimisation. However, they achieve speed at the expense of accuracy. Pencil beam algorithms perform reliably in homogeneous media and provide good agreement for depth–dose profiles, but struggle to accurately model lateral spread, tissue heterogeneities, and complex geometries. Due to this, Monte Carlo (MC) models such as Geant4 \citep{agostinelli2003geant4}, TOPAS \citep{perl2012topas}, GATE \citep{grevillot2020gate} and FLUKA \citep{bohlen2014fluka}, among others, are an alternative to analytical methods to perform the most accurate dose calculations, which can reduce the error margins by several millimetres \citep{paganetti2012range}. This is achieved by tracking each proton step by step and sampling each possible interaction according to its probability, which means this method tends to be much slower, compromising computational efficiency and therefore its suitability for routine treatment planning.

Optimised CPU codes such as MCsquare \citep{deng2020mcsquare} leverage multi-core architectures to achieve substantial reductions in computation time. To date, a number of highly optimised GPU-based Monte Carlo frameworks have demonstrated that near real-time proton dose calculation is achievable in practice, including FRED \citep{schiavi2017fred}, pGPUMCD \citep{maneval2019pgpumcd} and gPMC \citep{jia2012gpu}. Other GPU-based implementations such as MOQUI \citep{lee2022moqui} enable highly efficient dose scoring, while other approaches avoid simulating secondary particles to fully exploit GPU computing power \citep{shan2022virtual}. These implementations exploit massive parallelism to dramatically reduce wall-clock time. These methods are increasingly being adopted in research and clinical workflows \citep{feng2022gpu}.

The present work does not aim to compete with or replace these GPU-based Monte Carlo solutions in terms of absolute runtime. Instead, the focus here is on demonstrating that, proton-for-proton and for a given level of dosimetric accuracy, the proposed SDE formulation can reduce the computational effort required per proton to obtain a dose estimate compared to conventional particle-tracking Monte Carlo. In this sense, the SDE framework should be viewed as a complementary modelling paradigm: it preserves the stochastic physics of transport while replacing frequent small-scale interactions with an analytically motivated diffusion process, thereby improving efficiency at the level of the underlying physical model rather than through hardware acceleration alone. This provides a mathematically distinct and potentially synergistic alternative that, due to the inherent independence of proton paths in the SDE model, can be integrated with modern computing architectures in future implementations.

Rather than relying exclusively on increasing computational power, recent efforts have also focused on advancing the underlying physical and mathematical models to improve efficiency at the algorithmic level. In addition to variance reduction techniques such as the L$_{eq}$ formalism and related approaches \citep{maneval2018leq}, which accelerate Monte Carlo simulations by reducing statistical noise, there is growing interest in alternative formulations of the transport problem itself.

In this context, new mathematical approaches are needed for modelling proton beam treatment delivery, aiming to lie between high-fidelity Monte Carlo accuracy and pencil beam algorithm efficiency. \cite{Tristan2} proposed an approach based on optimisation by computing the fluence of the associated Boltzmann transport equation numerically. In \cite{SDE}, a complementary approach was introduced to solve the Boltzmann transport equation in backward form using a jump stochastic differential equation (SDE). The latter retains the probabilistic nature of Monte Carlo simulations while replacing frequent small-angle Coulomb scatterings with a physics-informed diffusion term. This reduces computational cost without sacrificing physical fidelity due to the central limit theorem. Moreover, \cite{KPP} showed that the two methods are consistent in the sense that they are alternative but compatible mathematical descriptions of the same underlying dynamics, expressed in terms of macroscopic fluence on the one hand and microscopic particles on the other. The reduced runtime of the proposed SDE approach compared to full Monte Carlo simulations makes it an attractive candidate for tasks such as plan verification, robustness evaluation, and adaptive planning, where high accuracy is required but computational cost remains a limiting factor.

In this article, we present an enhanced version of the SDE model with cross sections designed to match physical quantities. We compare its predictions to Geant4 as a reference standard using homogeneous and heterogeneous phantoms, and quantify both accuracy and computational speed. The two main objectives of this work are:
\begin{itemize}
\item To show SDE-based methods provide accurate dose predictions at low computational cost per proton in clinically relevant settings.
\item To highlight the inherent adaptability of the SDE framework, capable of accommodating a broad spectrum of model complexities, thus establishing a foundation for future optimisation based on clinical need.
\end{itemize}

\section{Methods}
\subsection{SDE-based transport formulation}
High-energy protons decelerating via interactions with subatomic particles are the basis of proton beam therapy. Deceleration is directly associated with energy deposition, and this energy transfer is a key component of attacking cancerous tissues. The rate of interaction of a proton with these subatomic particles increases as it loses energy, which leads to greater energy deposition towards the end of a proton track. Hence, the position at which a proton beam deposits the majority of its energy is controllable, limiting the exposure of surrounding healthy tissues. Indeed, the energy deposition can be arranged in such a way that there is effectively no exit dose from a proton beam directed towards tumour tissues. Proton beam therapy is distinguished from photon therapy by these phenomena. In \cite{SDE}, the basic principles of nuclear physics were used to model the dynamics of protons travelling through matter in three-dimensional space. The principal mathematical tool was a so-called stochastic differential equation (SDE) with jumps. The SDE method lends itself well to new Monte Carlo approaches which favourably balance accuracy and computational cost. In this section, we begin with a brief reminder of proton beam physics and how it plays into the modelling choices that lead to the SDE formulation, which is then described in detail. For a more detailed read on the background physics of proton beams, the reader is referred to \cite{NZ, Harald} and \cite{Gott}, among many possible sources.

\subsubsection{Physical basis and scope} \label{sec:proton_interactions}

We introduce the space-direction-energy phase space as a state space on which to describe proton beam dynamics. Let $D \subset \mathbb R^3$ be a closed, bounded and convex spatial domain, $\mathbb S_2$ be
the unit sphere in $\mathbb R^3$, and $\mathcal{E} = [\mathtt{e}_\mathtt{min},
\mathtt{e}_\mathtt{max}]\subset[0,\infty)$ be the interval of energies a proton can take. We define the energy-position-direction phase space as
$
  \mathcal{C} = \mathcal{E}\times D\times \mathbb{S}_2 .
$
 
To describe the dynamics of a proton at configuration $x= (\e, r,\omega)\in \mathcal{C}$, we consider transport as well as three classes of interactions with surrounding matter (See Figure \ref{fig:atom}). For further details, we refer the interested reader to \cite{NZ}. The three classes are:
\begin{itemize}
\item{\bf Transport.} The proton moves in a straight line in the direction $\omega$ in the absence of other interactions.

\item{\bf Inelastic Coulomb interaction.} We assume that a proton loses energy continuously via frequent collisions with orbital electrons. For an instantaneous configuration $x = (\e, r,\omega)$, there are two contributing factors to this continuous energy loss. Deterministic loss occurs at a rate given by the {\it stopping cross section}, $\varsigma_1(x)$, per unit track length. We model variability in the number of orbital electron collisions per unit track length by an additional stochastic loss term, governed by a Brownian motion whose volatility is given by the {\it energy-straggling cross section} $\varsigma_2(x)$. We assume that these collisions have no effect on the direction of motion of the relatively massive proton, or equivalently, that any effect is subsumed into the elastic scatter term described below. 
  \smallskip

  \begin{figure}[!ht]
  \centering
    \begin{tikzpicture}[scale=0.7]
        \nucleus
        \electron{1.5}{0.75}{80}
        \electron{1.2}{1.4}{260}
        \electron{4}{2}{30}
        \electron{4}{3}{180}
        \protoncollision{-6.}{0.}{160}
        \inelastic{-6.}{-2.}
        \elastic{-6.}{2.}
    \end{tikzpicture}
    \caption{The three main interactions of a proton with its surrounding matter. An elastic scattering (top) with a nucleus, a proton-nucleus collision which may be elastic or inelastic (centre), and an inelastic Coulomb interaction with an orbital electron (bottom).
      \label{fig:atom}
    }
    \vspace{-10pt}
\end{figure}

\item{\bf  Elastic scatter.} This class of events corresponds to changes in a proton's direction due to an interaction with a nearby nucleus whilst conserving energy between the proton and the nucleus. The three types of elastic scattering events we consider are as follows:

\noindent{\textit{Elastic Coulomb scatter:}} The proton passes close enough to a nucleus to feel a repulsive positive charge, resulting in a change in its direction of motion. We distinguish between large and small scatters as separate cases. For incoming direction \(\omega\) and outgoing direction \(\omega'\), let \(\theta\in[0,\pi]\) denote the polar scattering angle. We fix a small cut-off $\delta > 0$, such that $\theta > \delta$ is a large (elastic Coulomb) scatter while $\theta \le \delta$ is a small scatter.
A cross section $\sigma_{\rm e}(x) \pi_{{\rm e}}(x; \dd \omega' )$ specifies large scatter events, where $\sigma_{\rm e}(x)$ is the rate at which a proton $x = (\epsilon, r, \omega)\in\mathcal{C}$ undergoes a large elastic Coulomb scatter, while the density $\pi_{{\rm e}}(x; \dd \omega' )$ describes the distribution of the resulting outgoing direction $\omega'$.
Small scatter events are subsumed into an aggregated term, described below.
It is common to think of $\sigma_{\rm e}(x)$ as a rate per unit track length,  capturing the number nuclear interactions a proton has with its surrounding medium. 
\smallskip

\noindent {\textit{Elastic proton-nucleus collision and scatter:}} In rare cases, the incoming proton is absorbed into a nucleus, which becomes excited and releases another proton with the same energy as the incoming one. 
Since protons are indistinguishable, we model them identically to the elastic Coulomb scatters described above, so that $\sigma_e$ and $\pi_e$ are the cross section and probability density of both event types combined.
\smallskip

\noindent{\textit{Coulomb-nuclear interference:}} It is insufficient to treat the effects of elastic Coulomb scattering and elastic proton-nucleus collisions independently \citep{Nekrasov_2024}. The combination of these effects results in an additional perturbation in the outgoing angle due to so-called Coulomb-nuclear interference. We take the same cut-off $\delta$ to distinguish between small and large scatters, where large scatters are subsumed into the elastic Coulomb scatter and small scatters subsumed into the aggregation of small scatter, described below.
Mathematically, $\sigma_e$ and $\pi_e$ incorporate Coulomb-nuclear interference as well.
\smallskip

\item{\bf Aggregated small scatter.}
We model the net effect of all small scatter events via a diffusive component in the direction of the motion $\omega$ of a proton $x = (\epsilon, r, \omega) \in \mathcal{C}$ \citep[Section 3.9.6]{Vassiliev}. More precisely, the direction of transport $\omega$ undergoes a Brownian motion on $\mathbb{S}_2$ with state-dependent volatility $m(x) \geq 0$. We assume no energy loss is associated with these small scattering events.

\item{\bf Inelastic proton-nucleus collision and scatter.}  
When a proton interacts with a nucleus, it can undergo an inelastic collision, forming an excited nuclear state. This excited nucleus may subsequently emit a spray of secondary particles, transferring some of its energy to each particle. For the SDE model considered in this article, we assume that each collision will emit exactly one secondary proton which carries the majority of the outgoing energy and has outgoing direction $\omega'$. The energy deficit between the incoming and outgoing protons is accounted for by the recoil of the nucleus, other secondary particle emission which our SDE neglects, and the binding energy of the interaction. Nuclear inelastic scattering occurs with cross section, or rate per unit track length $\sigma_{{\rm ne}}(x)$. At an inelastic scattering event, the configuration $x = (\epsilon, r, \omega)$ of the incoming proton is transformed to the outgoing configuration $x' = (\e (1-u), r, \omega')$ 
with probability density $\pi_{{\rm ne}}(x; \dd \omega', \dd u)$, where
 $u\in(0,1]$.

\end{itemize}
 %As will be made evident in Section \ref{sec:results}, even with this secondary particle simplification, the accuracy of our model performs well against current Monte-Carlo codes without invoking the computational overhead incurred by these additional secondary tracks. Moreover, we emphasise that the SDE framework allows for multiple secondary particles, thus facilitating a hierarchical collection of models that trade computational cost for further accuracy. The result of this extension is a branching SDE, where SDEs of the form \eqref{eq:VSDE} are fitted to each particle type. However, for the scope of this article, we focus on how to fit a single type SDE for protons and leave this extension for future research.

Together, these interactions manifest in what is known as the Bragg peak: a graphical representation of the energy deposition per unit length along the axial direction of a proton beam. A typical Bragg peak curve is proportional to the average energy deposition per unit length along sequential proton tracks. Its characteristic shape shows a gradual increase in energy deposition followed by a sharp rise near the end of the proton range, and then a rapid fall-off. One of the defining features of the Bragg peak is its range, defined as the depth at which the dose falls to a predefined fraction of its maximum. Quantities such as R90 and R50, defined as the depths at which the dose falls to 90\% and 50\% of the maximum, respectively, are commonly used in the literature. In the context of the present work, the notion of a sequential proton track was introduced in \cite{SDE} to describe the configuration-space dynamics of a proton, concatenated with subsequent protons that continue its trajectory after proton-nuclear interactions, ultimately to absorption or a de-energised state.  Each sequential proton track is a random trajectory and a proton beam is comprised of many such tracks, all superimposed. We assume all sequential proton tracks are independent and hence, by the Law of Large Numbers, a proton beam is the average effect a sequential proton track multiplied by the number of protons in the beam.  

\subsubsection{Sequential proton track SDE formulation}\label{sec:sde-model}

To model the evolution of a proton's trajectory through the configuration space $\mathcal{C}$, we employ an enhanced version of the SDE first introduced in \cite{SDE}. We follow the usual physics and nuclear literature convention and index our SDE by the inherent `track length' of the sequential proton track which it describes.  
We define the dynamic evolution of the configuration variables along a sequential proton track via  
 $(Y_\ell, \ell \ge 0) = ( (\epsilon_\ell,  r_\ell, \omega_\ell), \ell\geq 0)$, where $\epsilon_\ell$ is the energy at track length $\ell\in\mathcal{E}$ of the sequential proton track,   $r_\ell\in D$ is the position of the sequential proton track at track length $\ell$, and $\omega_\ell\in\mathbb{S}_2$ is its direction of transport. The process  $Y$ depicts the stochastic evolution of a sequential proton track in configuration space ${\mathcal{C}}$. 

Since our focus is on simulating the sequential proton track, we introduce the evolution of $Y$ in terms of its Euler--Maruyama approximation: a standard numerical scheme for approximately simulating SDE paths. We will work on the lattice track lengths $(\ell = \Delta n, n\geq0)$, where $\Delta > 0$ is a small increment of path length. The collective evolution of $Y_n : = (E_n,   R_n, \Omega_n) := (\epsilon_{\Delta n},   r_{\Delta n}, \Omega_{\Delta n})$, for $n = 0,1,2,\dots$ is governed by  

\begin{equation}
\boxed{
\begin{array}{rl}
E_{n+1}& = (E_n -  \varsigma_1(Y_n)\Delta + \min\{\max\{\varsigma_2(Y_n)B_n, -\varsigma_1(Y_n)\Delta\}, \varsigma_1(Y_n)\Delta\})(1 - u_n)\\
&\\
R_{n+1} &= R_n +  \Omega_{n}\Delta   \\
&\\
\displaystyle\Omega_{n+1}& =
\Omega_n 
+ \Xi_n(m(Y_{n}))
+ D_n
\end{array}
\label{eq:VSDE}
}
\end{equation}
where:
\begin{itemize}
    \item $B_n$ in an increment of Brownian motion with variance $\Delta$ and the corresponding random term on the right-hand side has been truncated to ensure non-increasing energy while preserving the correct continuous energy loss.
    \item $\Xi_n(m(Y_{n}))$ is the time-$\Delta$ increment of spherical Brownian motion on $\mathbb{S}_2$ with volatility \(m(Y_{n})\), simulated via Algorithm 1 of \cite{Mijatovic}.
    It models the aggregate effect of small scattering events on the direction of the proton.
    \item $(u_n, D_n)$ takes value $(0, (0, 0, 0))$ with probability $1 - e^{- \Delta (\sigma_{\rm ne}(Y_n) + \sigma_{\rm e}(Y_n))}$, and otherwise is sampled from the mixture distribution 
    \begin{align}
\pi (Y_n; \dd \up', \dd u):= {}&
\frac{\sigma_{{\rm e}}(Y_n)}{\sigma_{\rm e}(Y_n) + \sigma_{\rm ne}(Y_n)}\pi_{{\rm e}}(Y_n; \dd \up' ) \delta_0(\dd u) \notag \\
&+\frac{\sigma_{{\rm ne}}(Y_n)}{\sigma_{\rm e}(Y_n) + \sigma_{\rm ne}(Y_n)}\pi_{{\rm ne}}(Y_n; \dd \up', \dd u).
\label{eq:Npi}
\end{align}
The former case corresponds to the absence of a large scattering event in the track length increment, while the latter corresponds to either an elastic ($u_n = 0$) or inelastic ($u_n \in (0, 1]$) scattering event.
\end{itemize}

The evolution in \eqref{eq:VSDE} continues until the proton exits the domain $D$, or its energy falls below the minimum threshold $\mathtt{e}_\mathtt{min}$.

In the following sections, we present our choices for the functionals $\sigma_{\rm e}, \pi_{\rm e}, \sigma_{\rm ne}, \pi_{\rm ne}, \varsigma_1, \varsigma_2$, and $m$. All are informed by physics so that $Y$ replicates the behaviour of sequential proton track.
This makes it possible to apply the SDE model to new materials simply by entering their chemical composition, mean excitation energy, and estimates of the nuclear scattering cross sections in ENDF format without the need to calibrate any free parameters. 

\subsubsection{Parameterisation for inelastic Coulomb scattering} \label{sec:inelastic_coulomb}
%In this section, we describe our choices of parameters for the SDE introduced in Section \ref{sec:sde-model}. We will use well-established physics-based models or, when such models are not available, numerical approximations fitted to experimental nuclear data. 

We recall from Section \ref{sec:sde-model} that energy losses due to inelastic Coulomb interactions are incorporated into the SDE \eqref{eq:VSDE} through two components: a deterministic stopping rate given by \(\varsigma_1\) and stochastic fluctuations (energy straggling) with volatility given by \(\varsigma_2\). To justify this model choice, we give a brief exposition on the theory of inelastic interactions (see \cite{BethandCorrections} for further details). First consider the setting of a proton with configuration $x = (E, R, \Omega)\in\mathcal{C}$ travelling through a homogeneous medium consisting of a unique element. Each inelastic interaction involves an energy transfer \(W\in [0,E]\) from the proton to the electrons of an atom in the medium. This energy transfer is completely characterized by the energy-dependent atomic energy-loss differential cross section \(f(W,E)\in \mathbb{R}^+\), \(W\in [0,E]\) and its moments
\begin{equation*}
    \sigma^{(n)}_E=\int_0^EW^nf(W,E)\mathrm{d}W, \quad n\geq 0.
\end{equation*}
If there are \(N\) atoms per \(\mathrm{cm}^3\), then \((N\sigma^{(0)}_E)\) is the mean free path length, \(\sigma^{(1)}_E/\sigma^{(0)}_E\) the mean energy loss in a collision, and \(\sigma^{(k)}_E/\sigma^{(0)}_E\), \(k\geq 2\), the \(k\)-th moment of the energy loss in a collision. Under idealised assumptions that \(N\) is sufficiently large and that interactions occur independently, the energy loss per unit track length is well-approximated by \(N\sigma^{(1)}_E+Z(N\sigma^{(2)}_E)^{1/2}\), where \(Z\sim\mathcal{N}(0,1)\). Indeed, this is just a consequence of the central limit theorem. Thus, for the SDE we choose \(\varsigma_1(x)=N\sigma^{(1)}_E\) and \(\varsigma_2(x)=(N\sigma^{(2)}_E)^{1/2}\). Explicit values of \(N\sigma^{(1)}_E\) and \(N\sigma^{(2)}_E\) are given by the well-established theory of the Bethe--Bloch formula with energy straggling.\\

\noindent \textbf{The Bethe--Bloch formula for mean energy loss.} The current state of the art model for mean energy loss from inelastic collisions is given by the Bethe--Bloch formula, along with its corrections. This quantum theory of stopping, originally due to Bethe, is based on the relativistic plane-wave Born approximation \citep{Bethesimple}. The Bethe--Bloch formula without corrections \citep[Section 3.2 and Appendix D]{Gott} per unit track length reads
\begin{equation}
\varsigma_1(x) = 0.3072 \frac{ Z \rho }{ A \beta^2 }\Bigg( \log\Big( \frac{ 2 m_e c^2 \beta^2 }{ I ( 1 - \beta^2 ) } \Big) - \beta^2 \Bigg) \; \frac{ \text{MeV} }{ \text{cm} },
\label{eq:bethbloch}
\end{equation}
with
\begin{equation*}
\beta^2 = \frac{ ( 2 m_p c^2 + E ) E }{ ( m_p c^2 + E )^2 }, 
\end{equation*}
where $\rho$ is the density of the medium in g/cm${}^3$, $Z$ is the atomic number of the medium, $A$ is the atomic mass of the medium, $m_p$ is the mass of a proton, $m_e$ is the mass of an electron, $c$ is the speed of light, and $I$ is the mean excitation energy of the medium. Importantly, this derivation assumes the idealised setting mentioned in the previous section. To account for deviations from the idealised setting, several additional correction terms exist which result in a more accurate model. The four correction terms to the Bethe--Bloch formula are known as the shell, density, Lindhard--Sørensen, and the Barkas correction term, respectively.
Their details can be found in equation Sections V--VII of \cite{BethandCorrections}. All are known to be negligible in the proton energy range of clinical interest \citep{BISCHEL1992345, MBAGWU2025112354, BethandCorrections, EnergyStragglingSalvat}, and hence we neglect them.

To extend \eqref{eq:bethbloch} to materials consisting of more than one element we use the so-called Bragg-additivity rule. This assumes that the material can be treated as a uniform mix of each of its constituting elements, and that interactions are independent between elements. In this case, the Bethe--Bloch formula reads \cite[Eq.\ (8)]{Gott}
\begin{equation}
\label{eq:BraggAditivty}
  \varsigma_1(x) = 0.3072 \sum_{i=1}^n\frac{ Z_i \rho_i }{ A_i \beta^2 }\Bigg( \log\Big( \frac{ 2 m_e c^2 \beta^2 }{ I ( 1 - \beta^2 ) } \Big) - \beta^2 \Bigg) \; \frac{ \text{MeV} }{ \text{cm} },
\end{equation}
where \(\rho_i,Z_i\) and \(A_i\) are the respective density, atomic number, and atomic mass of element $i$ in the medium.

Lastly, we note that due to the idealised setting used to derive the Bethe-Bloch formula, equation \eqref{eq:bethbloch} performs poorly when \(E\) is close to 0. Indeed, \eqref{eq:bethbloch} tends to \(-\infty\) as \(E\rightarrow 0\). To account for this problem, for the results presented in Section \ref{sec:results}, protons are absorbed upon reaching a kinetic energy of 0.05 MeV and their remaining energy is deposited at their current position. The choice of absorption energy threshold ws determined by numerical evaluation of or implementation of \eqref{eq:BraggAditivty} to ensure non-negative values, and is low enough that the remaining proton range is below 0.1 mm. Hence, the output of our simulations is unaffected for all practical purposes. An alternative approach using a logarithmic transform of the energy to guarantee non-negative energy for all track length has been considered in \cite{chronholm2025geometryenergysensitivitystochastic}.
\vskip 11pt
\noindent \textbf{Energy straggling.} The original derivation of \(N\sigma^{(2)}_E\) dates back to Bohr \cite[Eq.\ (3.4.5)]{BohrStraggling}. It neglects the binding of atomic electrons and assumes interactions with non-relativistic free electrons at rest. A more accurate formula based on the relativistic plane-wave Born approximation \citep{EnergyStragglingSalvat, jackson1975classical} reads
\begin{equation}
    \label{eq:energystraggling}
    \varsigma_2(x)^2=\frac{4\pi N_AZ\rho}{A}(\alpha \hbar c)^2\Bigg(\gamma^2\Bigg(1-\frac{\beta^2}{2}\Bigg)\Bigg)\; \frac{ \text{MeV}^2 }{ \text{cm} },
\end{equation}
where \(N_A\) is Avogadro's constant, \(\gamma\) is the Lorentz factor, $\alpha$ is the fine structure constant, $\hbar$ is the reduced Planck constant, and all other variables are as in \eqref{eq:bethbloch}. Identically to \eqref{eq:bethbloch}, due to the idealised setting assumed when deriving \eqref{eq:energystraggling}, an accurate formula for \(\varsigma_2\) requires additional correction terms. However, for the clinical energy range of protons with 0--150 MeV, the results of \cite{EnergyStragglingSalvat} show that they have negligible impact when the atomic number of the material is sufficiently small (less than 40), and between a 1--5\% impact for larger atomic numbers. Furthermore, since the effect of energy straggling is small in itself, these errors will be of an order of magnitude smaller when compared to the total energy loss.

Identically to \eqref{eq:BraggAditivty}, we extend \eqref{eq:energystraggling} to materials consisting of more than one element using the Bragg-additivity rule. This reads
\begin{equation*}
\varsigma_2(x)^2=\sum_{i=1}^n\frac{4\pi N_AZ_i\rho_i}{A_i}(\alpha \hbar c)^2\Bigg(\gamma^2\Bigg(1-\frac{\beta^2}{2}\Bigg)\Bigg)\; \frac{ \text{MeV}^2 }{ \text{cm} }.    
\end{equation*}

\subsubsection{Parameterisation for elastic scattering} \label{sec:elastic_scat}
We recall from Section \ref{sec:proton_interactions} that elastic scattering is divided into Coulomb and nuclear scattering events, and the interference between them. At small angles, the effect of elastic scattering is dominated by Coulomb contribution \cite[Section 6.2.7]{osti_1425114}. We recall that the SDE \eqref{eq:VSDE} models small angle elastic scattering through a diffusion on the sphere with volatility \(m\). Thus, we fit \(m\) using the state of the art for modelling Coulomb scattering, given by the theory of Moli\`ere \citep{Moliere}. The SDE \eqref{eq:VSDE} models large angle elastic scattering as a point process with rate \(\sigma_e\) and jump density \(\pi_e\). Since no satisfactory theory exists to describe the contributions from nuclear and nuclear-Coulomb elastic scattering, these parameters are fitted from the nuclear data libraries {\it ENDF/B-VIII.1} and {\it JEFF-4.0}. Both small and large angle contributions are described in detail below.
\vskip 11pt
\noindent \textbf{Moli\`ere's theory for small-angle elastic scattering.} 
The scattering density derived by Moli\`ere's has a Gaussian mode, but tails which are much heavier than Gaussian \citep{Gott}. Thus, we model small-angle elastic Coulomb scattering by taking a Gaussian approximation to Moli\'ere's distribution. Specifically, we use the Lynch--Dahl approximation to Moli\'ere multiple scattering \cite[Eq.\ (7)]{lynch/dahl:1991} which is given as follows. Consider a proton with configuration $x = (E,R,\Omega)\in\mathcal{C}$ travelling through a medium consisting of a single element. Relative to the initial direction of travel \(\Omega\), after a path length \(z\) cm, the new direction of travel in spherical coordinates has azimuthal angle uniform in \([0,2\pi]\) and energy-dependent polar angle that is the absolute value of a Gaussian with mean 0 and standard deviation \(\sigma(E)\) given by:
\begin{align*}
\chi_c^2 &= \frac{0.157 Z (Z + 1) \rho z}{A (p \beta)^2}, \\
\chi_{\alpha}^2 &= 2.007 \times 10^{-5} Z^{2/3} \frac{1 + 3.34 (Z \alpha /\beta)^2}{p^2},\\
w &= \frac{\chi_c^2}{2 \chi_{\alpha}^2 (1 - F)},\\
\sigma(E) &= \sqrt{\frac{\chi_c^2}{1 + F^2} \Big( \frac{1+ w}{w} \log(1 + w) - 1 \Big)},
\end{align*}
where \(p\) is the momentum in \(\text{MeV}/c\), and all other parameters are as in Section \ref{sec:inelastic_coulomb}. Note that the corresponding formula of \cite{lynch/dahl:1991} is missing the square root on the final line.
Here, $F \in (0, 1)$ is a truncation parameter describing the central fraction of the Coulomb scattering distribution taken into account. This is necessary because the tails of the distribution are not integrable, and thus we take $F = 0.98$.
This approximation is accurate, and advantageous because it does not use the radiation length of the medium, which is difficult to measure precisely.

For compounds, an effective $\chi_c^2$ is obtained by adding up all single-atom contributions, while an effective $\chi_{\alpha}^2$ is obtained via
\begin{equation*}
\log( \chi_{\alpha}^2 ) = \frac{\sum_i \frac{Z_i (Z_i + 1)}{A_i} \log(\chi_{\alpha, i}^2)}{\sum_i \frac{Z_i (Z_i + 1)}{A_i}}.
\end{equation*}
To convert these Gaussian updates into a rigorous mathematical object for use in the SDE, we use that for small path length step sizes \(z\) cm, the Lynch--Dahl approximation is well-approximated by the increment of a spherical Brownian motion of size \(z\) and volatility
\begin{equation*}
    m(x) = \frac{\sigma(E)}{z^{1/2}}.
\end{equation*}

For our simulations in Section \ref{sec:results}, we take \(z=0.05\) cm and use Algorithm 1 of \cite{Mijatovic} to compute fast and exact realisations of spherical Brownian motion. 

\noindent \textbf{Large angle elastic scattering via experimental data.} 
As shown in \cite{Gott}, the Lynch--Dahl approximation accurately models the effects of Coulomb elastic scattering for scattering angles up to \(\sim2.5\sigma(E)\). Beyond that, the light Gaussian tails fall too quickly to match the heavier tails of Moli\`ere's distribution. Computing these heavier tails requires an expensive root finding step \cite[Eq.\ (19)]{Gott}. Moli\`ere's theory also gives no insight into the contributions from nuclear and nuclear-Coulomb effects. Thus, we turn  instead to nuclear data libraries to model elastic scattering angles above \(2.5\sigma(E)\). Also, we note that although \(2.5\sigma(E)\) is energy-dependent, we fix \(2.5\sigma(E)=0.02\) radians as a cut-off throughout our simulations in Section \ref{sec:results}. This is done for the sake of computational efficiency, since a change in \(2.5\sigma(E)\) requires a re-evaluation of the nuclear data. Furthermore, for energies above 10 MeV, 0.02 radians is a good approximation of \(2.5\sigma(E)\) for {elements seen in proton beam therapy}.

We primarily use the {\it ENDF/B-VIII.1} data library, and refer to {\it JEFF-4.0} for elements not included in {\it ENDF/B-VIII.1}. For all elements except hydrogen, the scattering cross sections are given by {\it LAW=5} and {\it LTP=12} \cite[Section 6.2.7]{osti_1425114}. Consider a proton with configuration $x = (E,R,\Omega)\in\mathcal{C}$ travelling through a medium consisting of a single element. This format expresses the scattering cross section as
\begin{equation}
    \sigma^c_e(\mu,E) = \sigma_{R}^c(\mu,E) +\sigma_{NI}^c(E)P_{NI}(\mu,E) \text{ barns/sr}, \label{eq:elastic_large_barns}
\end{equation}
where \(\mu\) is the cosine of the polar scattering angle, \(\sigma_{R}^c(\mu,E)\) is the differential Coulomb scattering cross section in the {\it center-of-mass} frame given by Rutherford's formula with electronic screening ignored (conversion to the {\it lab} frame is given in Appendix \ref{appendix:angle_conv}), which reads
\begin{equation*}
    \sigma_{R}^c(\mu,E) = \frac{\eta^2}{k^2(1-\mu)^2},
\end{equation*}
where
\begin{equation*}
    k=\frac{A}{1+A} \sqrt{\frac{2m_pE }{\hbar^2c^2} } \times 10^{-26}, \quad \eta=Z \sqrt{\frac{\alpha^2 m_p}{2E}}\times 10^{6},
\end{equation*}
with parameters as in Section \ref{sec:inelastic_coulomb}, and \(\sigma_{NI}^c,P_{NI}\)  given by experimental data to account for nuclear and nuclear-Coulomb effects. Transforming \eqref{eq:elastic_large_barns} to a scattering rate per cm of path length travelled gives 
\begin{equation}
    \sigma_e(E) = \frac{2 N_A\pi\rho}{A}\int_{-1}^{\cos(2.5\sigma_E)} (\sigma_{R}^c(\mu,E) +\sigma_{NI}^c(E)P_{NI}(\mu,E))\mathrm{d}\mu \times10^{-24}. \label{eq: large_elastic_cm}
\end{equation}
The integral of \(\sigma_{R}^c(\mu,E)\) is calculated explicitly. The integral of \(P_{NI}(\mu,E)\) is calculated numerically by first fitting a cubic B-spline for \(P_{NI}(\mu,E)\), then integrating the resulting spline. The density of the scattering angle is given by 
\begin{equation}
    \pi_e(E;\mu) = \frac{(\sigma_{R}^c(\mu,E) +\sigma_{NI}^c(E)P_{NI}(\mu,E))}{\int_{-1}^{\cos(2.5\sigma_E)} (\sigma_{R}^c(\mu,E) +\sigma_{NI}^c(E)P_{NI}(\mu,E))\mathrm{d}\mu}, \quad \mu \in [-1,\cos(2.5\sigma(E))].
    \label{eq:large_elastic_pdf}
\end{equation}
To simulate a realisation from this density, it is sufficient to know its cdf which is given by
\begin{equation}
    \Pi_e(E;\nu) = \frac{\int_{-1}^{\nu}(\sigma_{R}^c(\mu,E) +\sigma_{NI}^c(E)P_{NI}(\mu,E))\mathrm{d}\mu}{\int_{-1}^{\cos(2.5\sigma_E)} (\sigma_{R}^c(\mu,E) +\sigma_{NI}^c(E)P_{NI}(\mu,E))\mathrm{d}\mu}, \quad \nu \in [-1,\cos(2.5\sigma(E))], \label{eq:large_elastic_cdf}
\end{equation} 
where the numerator is computed identically to \eqref{eq: large_elastic_cm}. For the simulations presented in Section \ref{sec:results}, both \eqref{eq: large_elastic_cm} and \eqref{eq:large_elastic_cdf} are computed for a discrete lattice of energy values and linear interpolation is used to extend to the full energy range. Once a scattering angle is simulated, it is converted into the {\it lab} frame using \eqref{eq:cm_to_lab}.

For hydrogen, the scattering cross sections are given by {\it LAW=5} and {\it LTP=1}. This format expresses the scattering cross section as 
\begin{align}
\sigma_{e}^c(\mu, E)=\sigma_{R}^c(\mu, E) & + \sigma_{N}^c(\mu,E)\text{ barns/sr},
\end{align}
where since the incident and target particles are identical, Rutherford's formula with electronic screening ignored now reads
\begin{equation*}
  \sigma_{R}^c(\mu, E)=  \frac{2 \eta^2}{k^2\left(1-\mu^2\right)}\left[\frac{1+\mu^2}{1-\mu^2}-\frac{1}{2} \cos \left(\eta \ln \frac{1+\mu}{1-\mu}\right)\right],
\end{equation*} 
furthermore 
\begin{align}
     \sigma_{N}^c(\mu,E) =&\sum_{l=0}^{\mathrm{NL}} \frac{4 l+1}{2} b_l(E) P_{2 l}(\mu) \nonumber\\
     &-\frac{2 \eta}{1-\mu^2} \operatorname{Re}\left\{\sum_{l=0}^{\mathrm{NL}}\left[\begin{array}{r}
(1+\mu) \exp \left(i \eta \ln \frac{1-\mu}{2}\right) \\
+(-1)^l(1-\mu) \exp \left(i \eta \ln \frac{1+\mu}{2}\right) \label{eq:hydrogen_not_ruther}
\end{array}\right] \frac{2 l+1}{2} a_l(E) P_l(\mu)\right\}, 
\end{align}
where \(P_{l}\) is the Legendre polynomial of degree \(l\), \(a_l(E)\) and \(b_l(E)\) are respectively complex and real coefficients derived from experimental data, and the value of NL represents the highest partial wave contributing to nuclear scattering which is equal to 7 for the hydrogen dataset provided by the {\it ENDF/B-VIII.1} library. As before, transforming \eqref{eq:elastic_large_barns} to a scattering rate per cm of path length travelled gives
\begin{equation}
    \sigma_e(E) = \frac{2 N_A\pi\rho}{A}\int_{-1+\delta}^{\cos(2.5\sigma_E)} (\sigma_{R}^c(\mu,E) +\sigma_N^c(\mu,E))\mathrm{d}\mu \times10^{-24}, \label{eq:large_elastic_cm_hydrogen}
\end{equation}
where \(\delta>0\) is a user-defined truncation parameter required since the tail of the Rutherford cross section for identical particles as \(\mu\rightarrow -1\) is not integrable (for the simulations presented in Section \ref{sec:results} we take \(\delta=1+\cos(\pi-0.2)\)). Similarly, accounting for \(\delta\) in the lower limit, the associated density and cdf of the scattering angle are given by \eqref{eq:large_elastic_pdf} and \eqref{eq:large_elastic_cdf} respectively. The coefficients \(a_l(E),b_l(E)\) are given for a set of lattice points which we extend using a cubic B-spline. Moreover, for given coefficients \(a_l(E),b_l(E)\), the integrals in \eqref{eq:large_elastic_cdf} can be computed explicitly as shown in Appendix \ref{appendix:angle_conv}. Finally, for compound materials, we again assume Bragg-additivity meaning the point process \(\sigma_e,\pi_e\) can be decomposed as the sum of \(n\) independent point processes, where process \(i\) corresponds to the elastic scattering due to atoms of element \(i\). 

\subsubsection{Parameterisation for inelastic proton-nucleus interactions}
As before, consider a proton with configuration \(x = (E,R,\Omega)\in \mathcal{C}\) travelling through a medium consisting of a single element. We recall from \eqref{eq:VSDE} that inelastic proton-nucleus interactions occur in the SDE at an element-dependent rate \(\sigma_{ne}\), and the outgoing energy and angle of the proton after such an interaction is given by the density \(\pi_{ne}\). There exists no satisfactory theory to describe the effects of these interactions, so we use experimental data to model \(\sigma_{ne}\) and \(\pi_{ne}\). As for large elastic scattering, we use the \textit{ENDF/B-VIII.1} and \textit{JEFF-4.0} nuclear data libraries. These libraries use the Kalbach-Mann systematics representation to model inelastic proton-nucleus interactions \citep{Kalbach_systematics} which is given as follows. The scattering rate per cm of path length travelled, \(\sigma_{ne}(E)\), is given for a lattice of points in \(\mathcal{E}\) which we fit to a cubic B-spline to extend to \(\mathcal{E}\). For \(E \in \mathcal{E}\), \(u \in [0,1]\), let \(\pi_{ne}(E;\mathrm{d}u)\) denote the marginal distribution of the outgoing energy given incident energy \(E\). For a given set of lattice points \(E_i\in \mathcal{E}\), \(U_j \in [0,1]\), \(\pi_{ne}(E_i;U_j)\) is given along with a fitted parameter for the outgoing angle density, denoted \(r(E_i,U_j)\in[0,1]\). We extend these to \(\mathcal{E}\) and \([0,1]\) respectively using linear interpolation. Finally, the distribution of the cosine of the polar scattering angle in the \textit{center-of-mass} frame (conversion to the {\it lab} frame is given in Appendix \ref{appendix:angle_conv}) given incident energy \(E\) and outgoing energy \((1-u)E\) is given by
\begin{equation*}
    \pi_{ne}(\mu|E,u) = \frac{a(E,u)}{2\sinh(a(E,u))}\left(\cosh(a(E,u)\mu)+r(E,u)\sinh(a(E,u)\mu)\right),
\end{equation*}
where \(a(E,u)\) is a known constant given in Section 6.2.3 of \cite{osti_1425114}. The cdf of this density is invertible, thus the outgoing angle, \(\mu\), can be simulated using
\begin{align*}
    &\mu = (C+(C^2-r(E,\mu)^2+1)^{1/2})/(r(E,\mu)+1), \\ &C=2\sinh(a(E,\mu))U+r(E,\mu)\cosh(a(E,\mu))-\sinh(a(E,\mu)),
\end{align*}
with \(U\sim \text{Unif}[0,1]\). The setting of compound materials is handled identically to large elastic scattering (see Section \ref{sec:elastic_scat}).

\subsection{Benchmark and comparison}
\subsubsection{Reference Monte Carlo simulations}
To verify the results from the proposed model, the Monte Carlo simulation toolkit Geant4 (version 11.3) \citep{agostinelli2003geant4} was used as the benchmark. Geant4 was selected for its flexibility, including the ability to enable or disable specific physical processes and customise physics lists. For this study, we have used the prebuilt \texttt{QGSP\_BIC\_EMZ} physics list, as it provides accurate proton transport modelling via the Binary Cascade model and includes the \texttt{EMZ} option, which offers the most precise electromagnetic physics list in Geant4, optimised for low-energy proton transport. 

\subsubsection{Test cases}
All simulations were configured with identical parameters and evaluated using three phantom geometries designed to probe different dosimetric effects. The first geometry is a homogeneous 20$\times$20$\times$20 cm$^3$ water phantom, used as a reference case for baseline dose and range in a uniform medium. The second is a slab phantom of the same size, composed of 2 cm water, 1 cm bone, 2 cm compressed lung, and a distal water region. The layers are laterally homogeneous and are intended to assess the impact of longitudinal heterogeneities on proton range and dose deposition. Slab thicknesses were chosen such that all materials are traversed by both 100 MeV and 150 MeV proton beams, while ensuring that the Bragg peak occurs within the final water layer. The final geometry is a water phantom of the same dimensions, containing a 2 cm thick bone insert located in the upper half of the phantom at a depth of 3 cm, designed to investigate the effect of lateral heterogeneities. These configurations are hereafter referred to as the water, slab, and insert phantoms, respectively. 

In the heterogeneous phantom simulations, the bone and lung regions were defined using standard material definitions. The bone insert was modelled using the Geant4 material \texttt{G4\_B-100\_BONE} with a density of 1.45 g/cm$^3$, while the lung region was modelled using the ICRP 110 reference human phantom composition with a density of 0.385 g/cm$^3$. These material definitions were used consistently in both the Geant4 reference simulations and the SDE model parameterisation.

For all cases, a monoenergetic proton beam (energy of 100 or 150 MeV, nozzle radius of 5 mm, energy spread $\sigma_E$ of 0.1 keV and radial dispersion $\sigma_r$ of 0.05 rad) was directed perpendicularly into water, with no air gap between the source and the phantom. The chemical compositions and densities for all materials are the same for Geant4 and the SDE model. For all cases, $1\times 10^6$ protons were fired per run, and dose was scored in both cases using a 3D grid of 1 mm$^3$ voxels across the irradiated volume, enabling pointwise comparison of the resulting dose distributions. The dose in each voxel was calculated as the deposited energy divided by the local voxel mass, corresponding to dose-to-medium. The SDE model requires angular limits for backscatter and Rutherford events, which were set at 0.04 and 0.02 rad, respectively. Since the SDE model employs a fixed step length, this was set to 0.5 mm for all simulations. For consistency, the maximum step length in Geant4 was constrained to the same value to ensure a fair comparison in terms of time performance. For both simulators, the energy deposition is assigned at the step midpoint to minimise bias in regions of steep gradients.

\subsubsection{Comparison metrics}
For quantitative assessment, integral dose-depth curves, central-axis depth-dose distributions, range calculations and lateral profiles were evaluated. In addition, a full 3D gamma analysis \citep{Gamma} and 3D voxel-to-voxel dose difference maps were used to quantify spatial agreement between the SDE and Geant4 dose distributions, where the SDE output is treated as the evaluated distribution, while the Geant4 distribution serves as the reference. 

For range estimation, the integrated depth–dose curve was linearly interpolated to determine the depth (to 0.1 mm resolution) at which the dose on the distal fall-off decreases to 90\% of its maximum (R90). Voxel-wise dose differences between the evaluated and reference distributions were computed within a reference-defined region of interest. The region of interest was defined as voxels receiving at least 1\% of the maximum reference dose. Differences were normalised to the 99th percentile of the reference dose, D99, computed within this region, in order to provide a near-maximum dose scale while reducing sensitivity to isolated high-dose outliers.

Gamma analysis was implemented using the \texttt{pymedphys} Python package, which incorporates methods presented in \cite{wendling2007fast}. Given that gamma analysis is employed here as a numerical similarity metric between two simulated distributions, rather than as a clinical QA criterion, the distance-to-agreement (DTA) parameter is not intended to compensate for setup or alignment uncertainties, and a stricter dose difference (DD) criterion is chosen. In this study, local dose normalisation was applied, with a dose threshold of 1\% of the maximum dose to suppress low-dose noise. Given the exact geometric correspondence and a voxel size of 1 mm, strict acceptance criteria of DD = 2\% and  DTA = 0.5 mm were used throughout. For context, clinical QA guidelines typically report $\geq$ 95\% pass rates using 3\%/2 mm criteria with a 10\% dose threshold under global normalisation \citep{miften2018tolerance}.

To reduce the influence of the discrete voxel grid, the gamma analysis algorithm in \texttt{pymedphys} employs on-the-fly linear interpolation between voxels, allowing $\gamma$ to be evaluated at sub-voxel positions. This process is controlled by the interpolation fraction, which specifies the interpolation step size as a fraction of the DTA criterion. In this study, the interpolation fraction was set to 10, corresponding to a step size of 0.2 mm. This interpolation does not alter the underlying dose distributions, but enables a more accurate determination of the gamma index.

In addition to dosimetric accuracy, computational performance was recorded as a reference metric. Wall-clock runtimes correspond to the elapsed execution time reported by the Unix \texttt{time} utility. All simulations were executed on a MacBook Pro equipped with an Apple M2 Max processor using $1\times10^6$ primary protons in single-thread mode. The reported timings are not intended as a direct performance comparison with clinically optimised GPU-based Monte Carlo engines, but to illustrate the computational cost of the proposed SDE formulation relative to a standard CPU-based Monte Carlo implementation under identical conditions. Due to its structure, the SDE solver exhibits a predictable per-particle execution cost and is well suited to parallel execution, making it a natural candidate for multi-threaded and GPU-based implementations.

\section{Results} \label{sec:results}

\subsection{Homogeneous water phantom}

\begin{figure}[!t]
    \centering
    \begin{subfigure}[t]{0.48\textwidth}
        \centering
        \includegraphics[width=\textwidth]{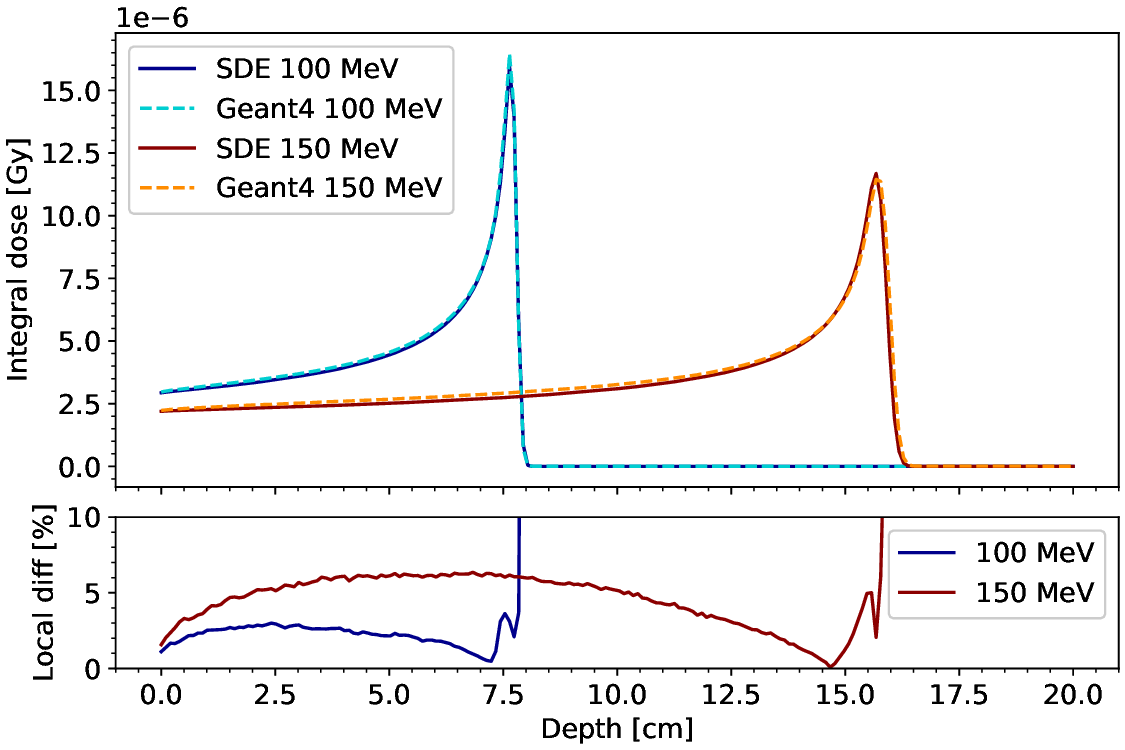}
        \caption{}
        \label{subfig:IDD_water}
    \end{subfigure}
    ~ 
    \begin{subfigure}[t]{0.48\textwidth}
        \centering
        \includegraphics[width=\textwidth]{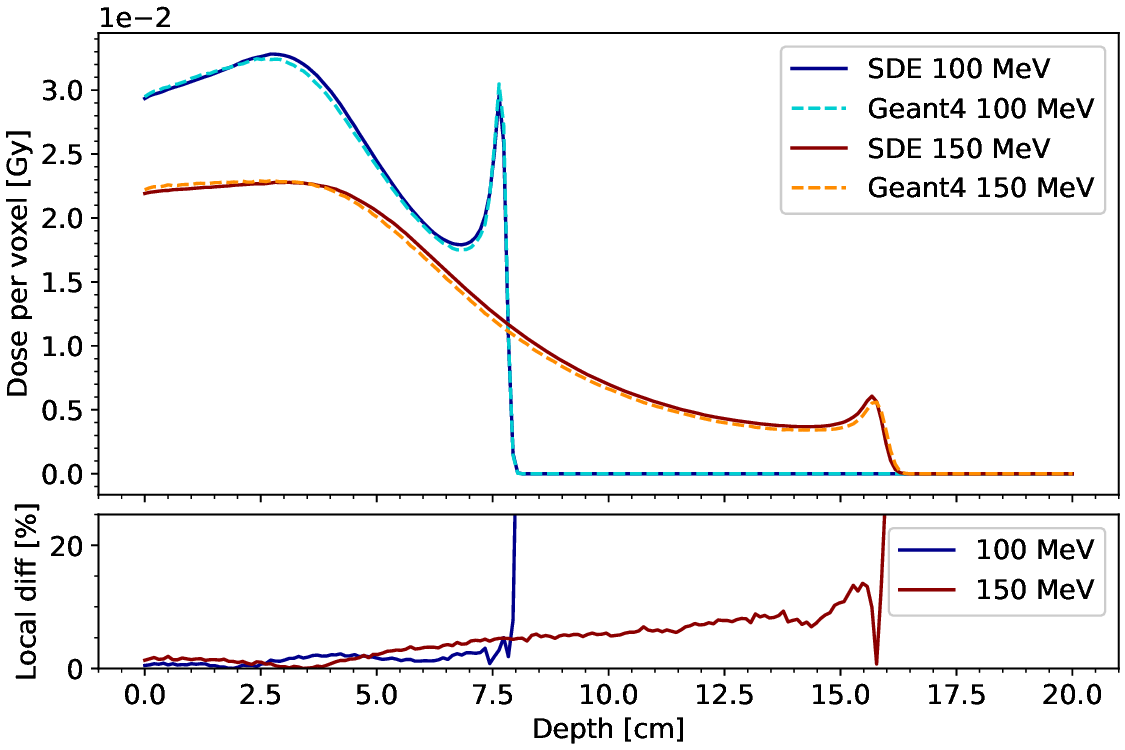}
        \caption{}
        \label{subfig:1Dslice_water}
    \end{subfigure}
    \caption{1D comparison between SDE and Geant4 for two monoenergetic proton beams (100 and 150 MeV) in a homogeneous water phantom, including pointwise calculations for percentage dose differences in the lower subplots. (a) Integral depth-dose curves, (b) Central-axis depth-dose curves.}
    \label{fig:1Dcomp_water}
\end{figure}

We present the dose comparison calculations for the baseline case of a water phantom being irradiated with a proton beam of 100 MeV and 150 MeV. This case serves as a foundation for the overall accuracy of the model. From the 3D dose arrays that are obtained in both reference and evaluated models, the integrated depth-dose curves were obtained and shown in Figure \ref{subfig:IDD_water}, while differential depth-dose curves were obtained from the central voxels of the 3D dose arrays for a more detailed comparison of the core beam behaviour are shown in Figure \ref{subfig:1Dslice_water}. 

Deviations between the integrated depth-dose curves are below 3.5\% for a 100 MeV beam and below 6.5\% for a 150 MeV when evaluated pointwise before the Bragg peak. Beyond the peak, two factors lead to systematic differences between the models. First, the SDE model applies an energy threshold below which proton transport is terminated, which causes the dose to fall off slightly faster than in Geant4. Second, because the SDE does not model dose contributions from secondary particles such as neutrons and gammas, it does not reproduce the remanent dose at remote locations predicted by Geant4. Together, these effects result in a steeper fall-off and zero dose deposition at greater depths in the SDE results. Nevertheless, the proton range is accurately reproduced, with R90 calculations agreeing within less than 0.1 mm for 100 MeV and within 0.6 mm for 150 MeV. This evidences the accuracy in the electromagnetic processes modelled in our approach. Moreover, the percentage differences in the integrated depth-dose curves are highest at mid-depths and decrease towards the Bragg peak. This behaviour indicates that the SDE model achieves its best agreement with Geant4 in the high-dose region near the Bragg peak, where dose accuracy is most critical. 

The central-axis depth dose curves are highly sensitive to local variations in scattering, energy loss and voxel sampling, which results in a slightly higher overall percentage difference compared to integrated depth-dose curves, but still remaining within 6\% for 100 MeV and 16\% for 150 MeV. Moreover, these curves are sensitive to the angular limits set in the SDE model for Rutherford and backscatter events. The agreement is highest at shallow depths, after which it starts decreasing. Overall, the trends suggests that the SDE model reproduces the core beam behaviour reliably, while minor discrepancies at depth are expected due to the differences in scattering modelling.

\begin{figure}[!ht]
    \centering
    \begin{subfigure}[t]{0.48\textwidth}
        \centering
        \includegraphics[width=\textwidth]{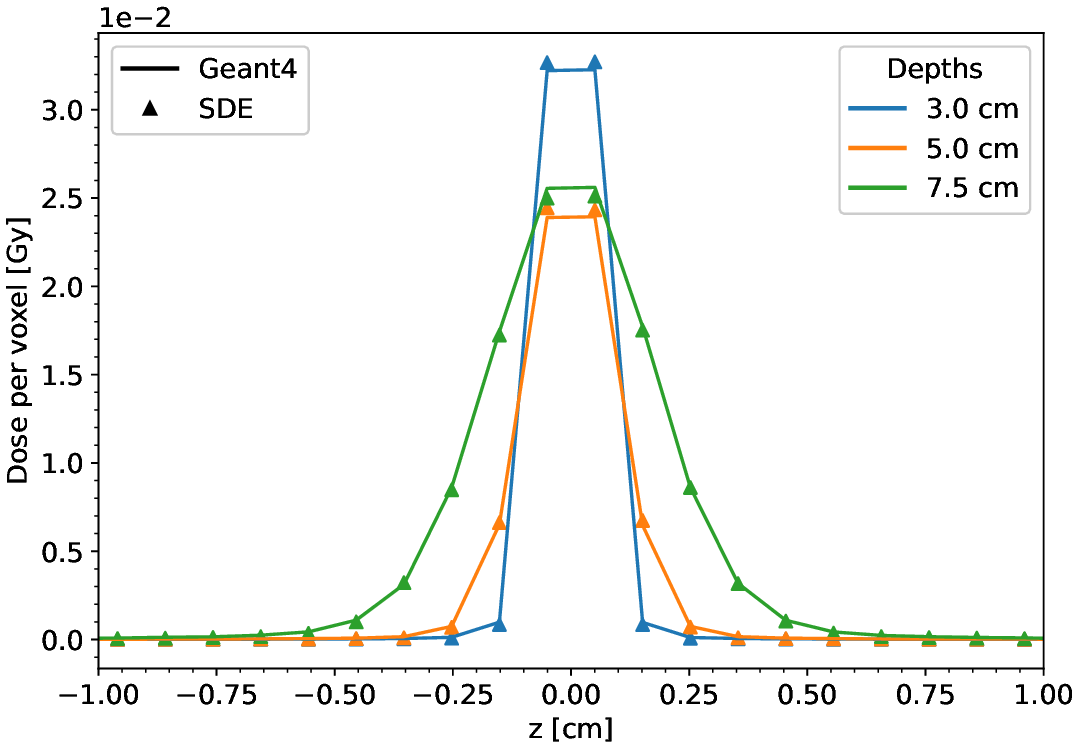}
        \caption{}
        \label{subfig:lateral100MeV_homogeneous}
    \end{subfigure}%
    ~ 
    \begin{subfigure}[t]{0.48\textwidth}
        \centering
        \includegraphics[width=\textwidth]{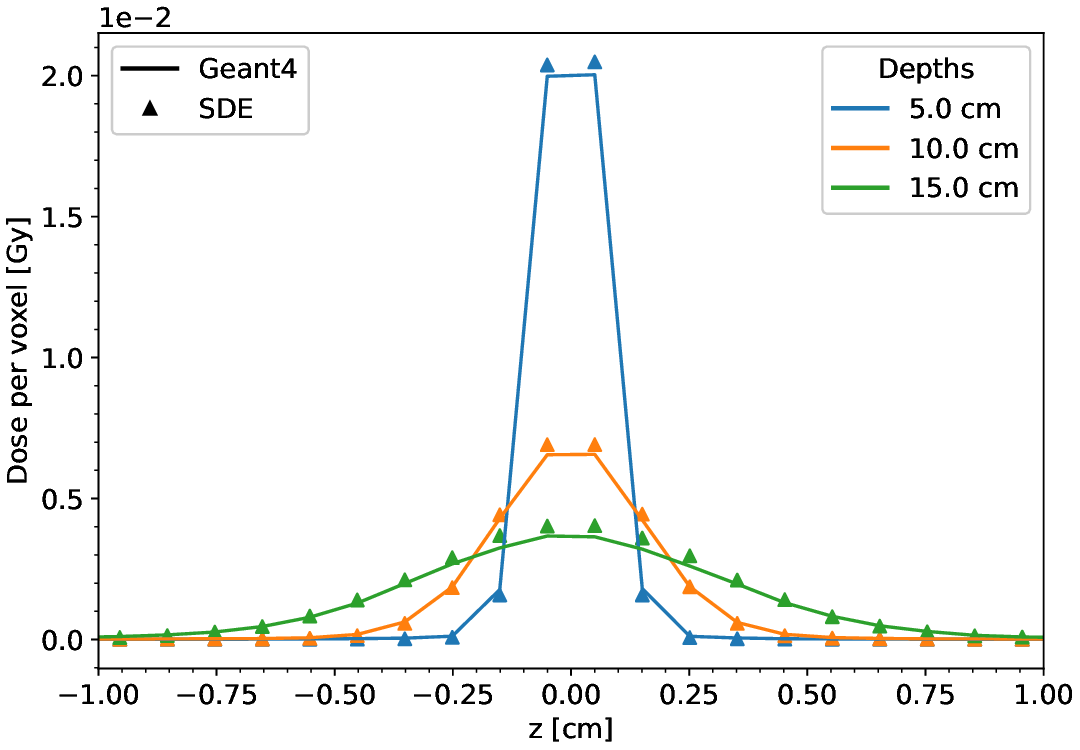}
        \caption{}
        \label{subfig:lateral150MeV_homogeneous}
    \end{subfigure}
    \caption{Lateral profile comparison between SDE (marker points) and Geant4 (solid lines) for two proton energies in a homogeneous water phantom (a) using a 100 MeV beam, (b) using a 150 MeV beam.}
    \label{fig:1Dprofiles}
\end{figure}

To further evaluate the beam's lateral spread and its consistency with depth, Figures \ref{subfig:lateral100MeV_homogeneous} and \ref{subfig:lateral150MeV_homogeneous} show lateral profiles at selected depths for 100 and 150 MeV proton beams, respectively. The SDE model consistently reproduces the overall Gaussian beam shape, showing similar widths to Geant4 across all depths. This indicates that the dominant multiple scattering behaviour of the beam is well captured. 

\begin{figure}[t!]
    \begin{subfigure}[t]{0.56\textwidth}
        \centering
        \includegraphics[width=\textwidth]{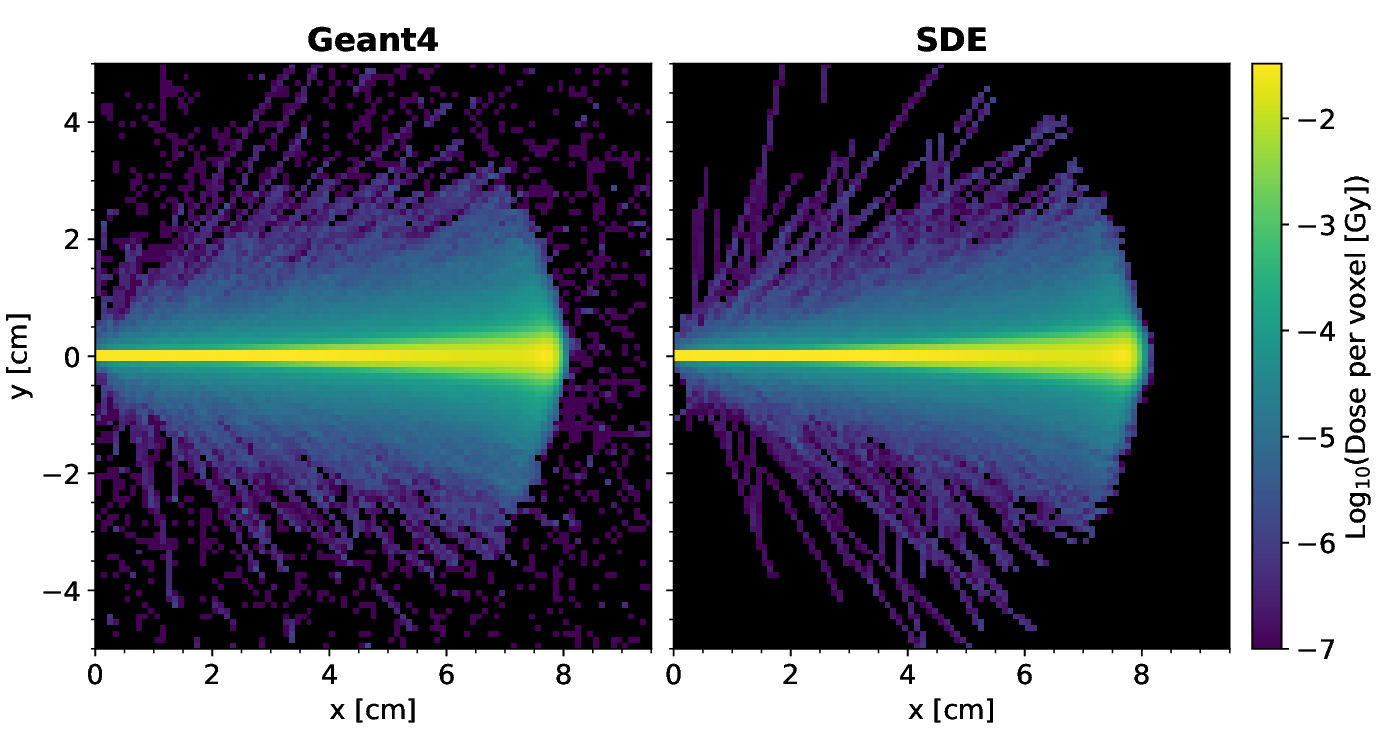}
        \caption{}
        \label{subfig:2Dslice100MeV_water}
    \end{subfigure}
    \begin{subfigure}[t]{0.43\textwidth}
        \centering
        \includegraphics[width=\textwidth]{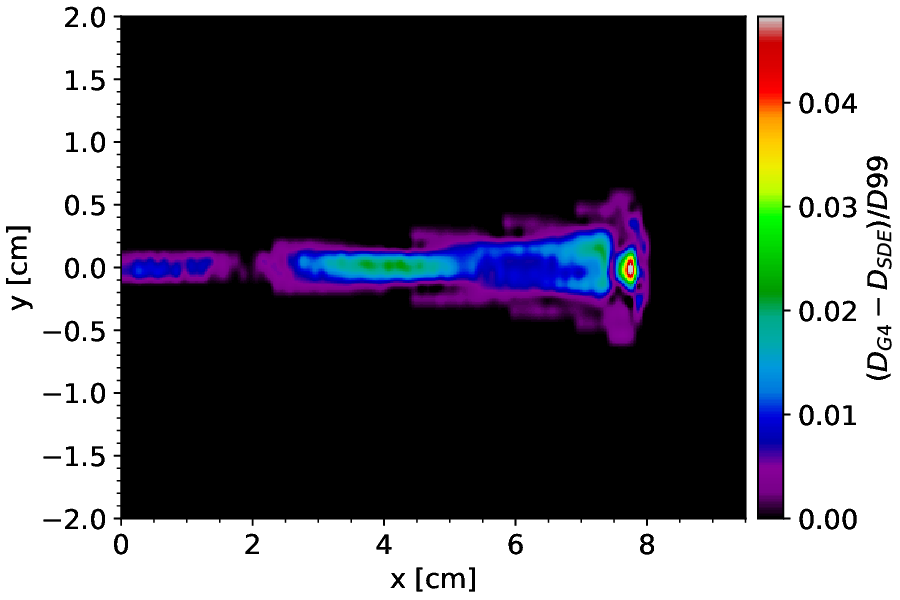}
        \caption{}
        \label{subfig:doseDiff100MeV_water}
    \end{subfigure}
    \vspace{-2mm}
    \begin{subfigure}[t]{0.56\textwidth}
        \centering
        \includegraphics[width=\textwidth]{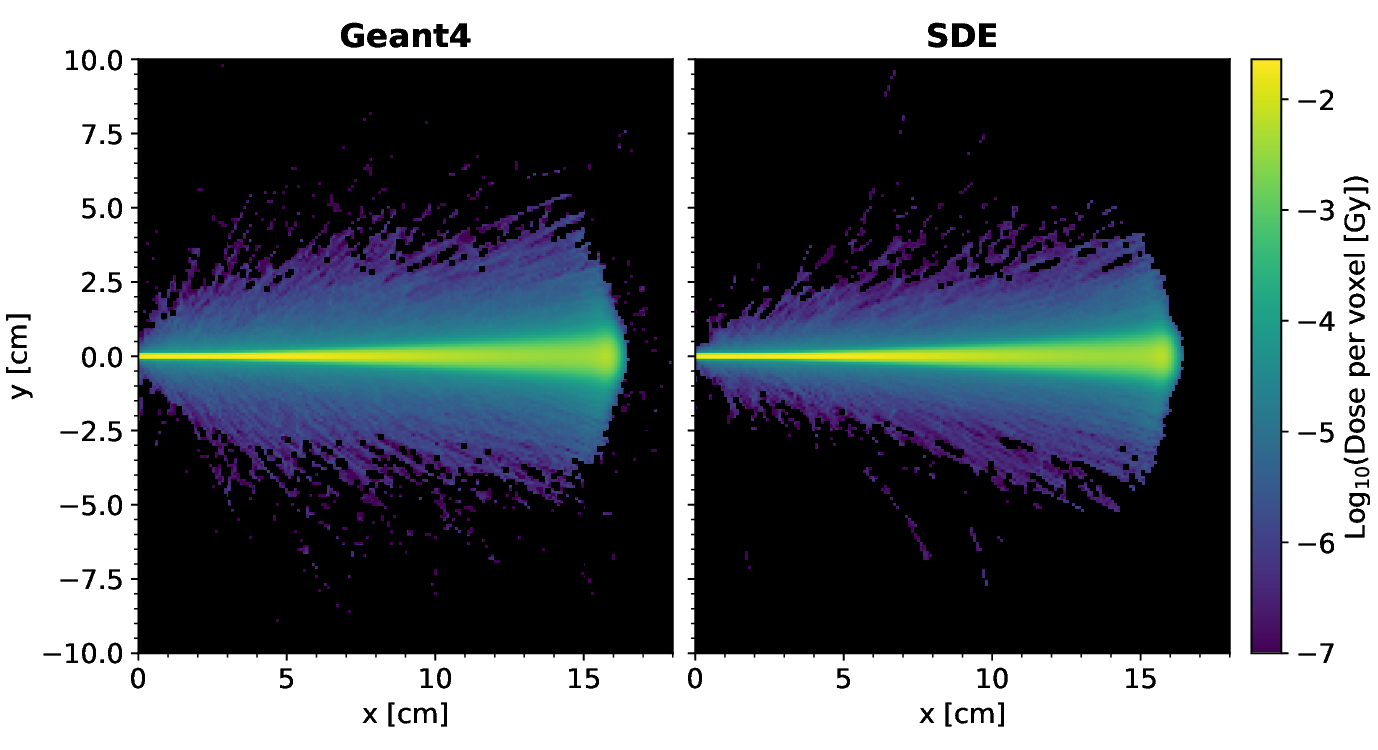}
        \caption{}
        \label{subfig:2Dslice150MeV_water}
    \end{subfigure}
    \hspace{2mm}
    \begin{subfigure}[t]{0.43\textwidth}
        \centering
        \includegraphics[width=\textwidth]{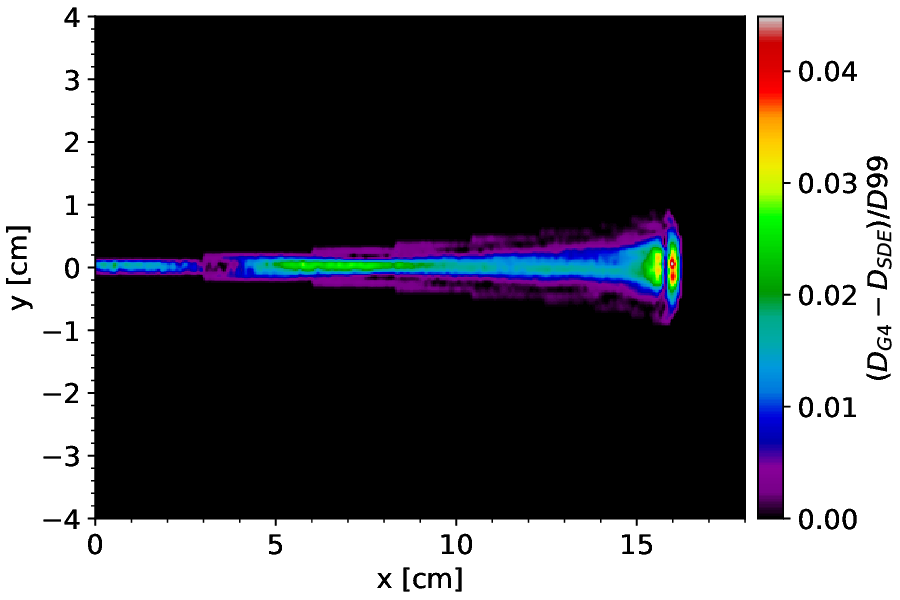}
        \caption{}
        \label{subfig:doseDiff150MeV_water}
    \end{subfigure}
    \caption{2D dose distribution comparison between the SDE model and Geant4 in a homogeneous water phantom for monoenergetic proton beams. (a) Central-axis dose distribution and (b) corresponding relative dose difference map for a 100 MeV beam, normalised to the 99th percentile (D99) of the Geant4 reference dose. (c,d) Same quantities for a 150 MeV beam.}
    \label{fig:2DsliceComp1}
\end{figure}

Small, systematic differences in the absolute dose are observed across the profiles. Although both the SDE model and Geant4 model scattering processes using partitioning between single-scattering and multiple scattering regimes using angular cutoffs, differences in how these regimes are implemented and coupled to energy deposition lead to modest variations in the proton path-length distribution and the associated dose. These effects exhibit increased sensitivity at lower beam energies, where scattering processes play a larger role. Despite these differences, the close agreement in profile widths demonstrates that the spatial dispersion of the beam is accurately reproduced. A more noticeable deviation is observed at the Bragg peak location for the 100 MeV beam ($x=7.5$ cm), where the peak dose predicted by our approach differs from Geant4. This reflects the sensitivity of the peak region to discretisation and the treatment of angular scattering.

In addition, a central-axis 2D slice was extracted to visualise the overall dose distributions, along with a voxel-wise difference map for the same slice. These results are shown in Figure \ref{fig:2DsliceComp1}. The dose maps using a 100 and 150 MeV are shown in Figures \ref{subfig:2Dslice100MeV_water} and \ref{subfig:2Dslice150MeV_water}, respectively, and illustrate the strong agreement between the SDE and Geant4, including in low-dose regions where the SDE accurately captures lateral spread. The main differences are observed in the lowest dose values, corresponding to non-local dose deposits present only in the Geant4 maps. These arise from secondary particles that interact with atoms at distant locations, producing small, remote dose contributions that the current SDE model does not simulate.

The relative dose difference maps in Figures \ref{subfig:doseDiff100MeV_water} and \ref{subfig:doseDiff150MeV_water} provide a spatially resolved, qualitative assessment of the agreement between the SDE model and Geant4. Since the differences are normalised to a global high-dose reference (D99 of the Geant4 dose), the reported values should not be interpreted as local voxel-wise dose errors. Instead, they indicate the magnitude and spatial distribution of discrepancies relative to the characteristic high-dose scale of the problem. Bearing this in mind, there is good agreement across the high-dose region, with the largest relative differences localised around the distal fall-off, where steep dose gradients amplify small discrepancies between the two models. In the entrance and plateau regions, differences remain low. The pattern and magnitudes of the difference maps is similar for both beam energies, suggesting consistent model behaviour over the energy range. 

For completeness and to illustrate the potential of our model for clinical purposes, gamma analysis was performed in 3D for thorough comparison for the two proton energies under study. Pass rates of 97.1\% for the 100 MeV case and 95.4\% for the 150 MeV case were obtained using stricter criteria than the conventional clinical ones (DD = 2\%/0.5 mm, 1\% dose threshold). This demonstrates that the SDE model achieves clinically acceptable agreement with Geant4. In terms of computational performance, the SDE model is 2.5 times faster than the single-threaded Geant4 model when using the 100 MeV beam, and 2.8 times faster using the 150 MeV beam.

\subsubsection{Comparison against different Geant4 physics lists}
\begin{figure}[h!]
    \centering
    \begin{subfigure}[t]{0.49\textwidth}
        \centering
        \includegraphics[width=\textwidth]{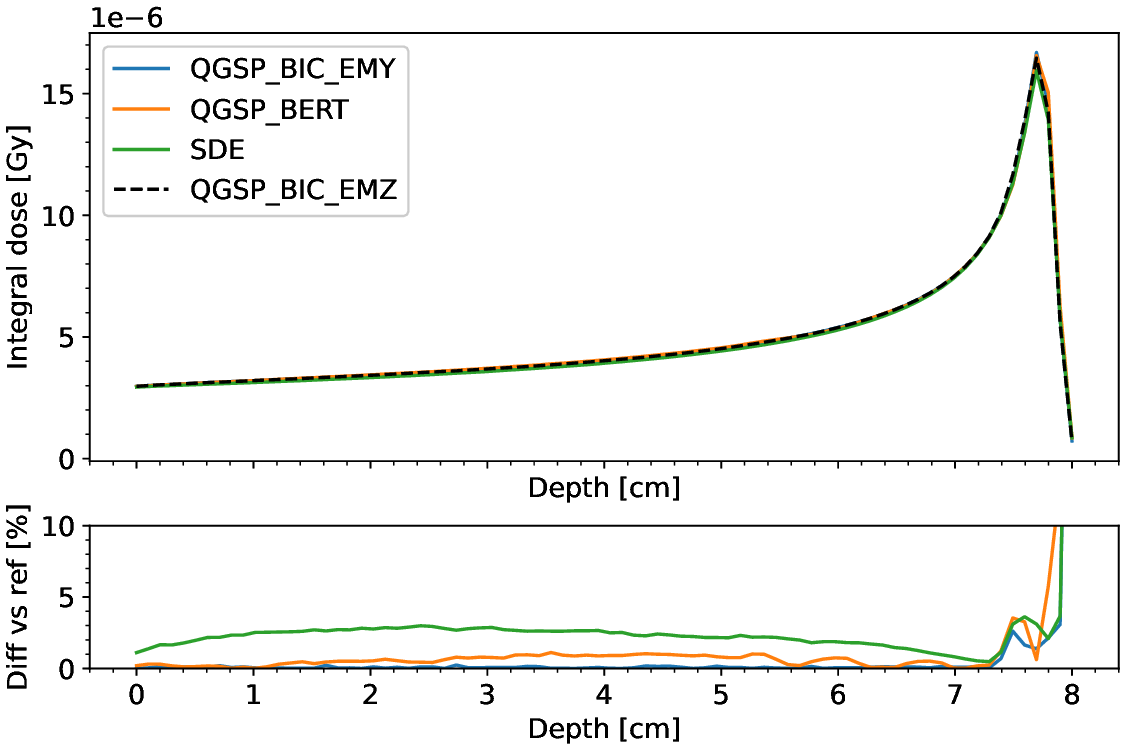}
        \caption{}
        \label{subfig:IDD3}
    \end{subfigure}%
    ~ 
    \begin{subfigure}[t]{0.49\textwidth}
        \centering
        \includegraphics[width=\textwidth]{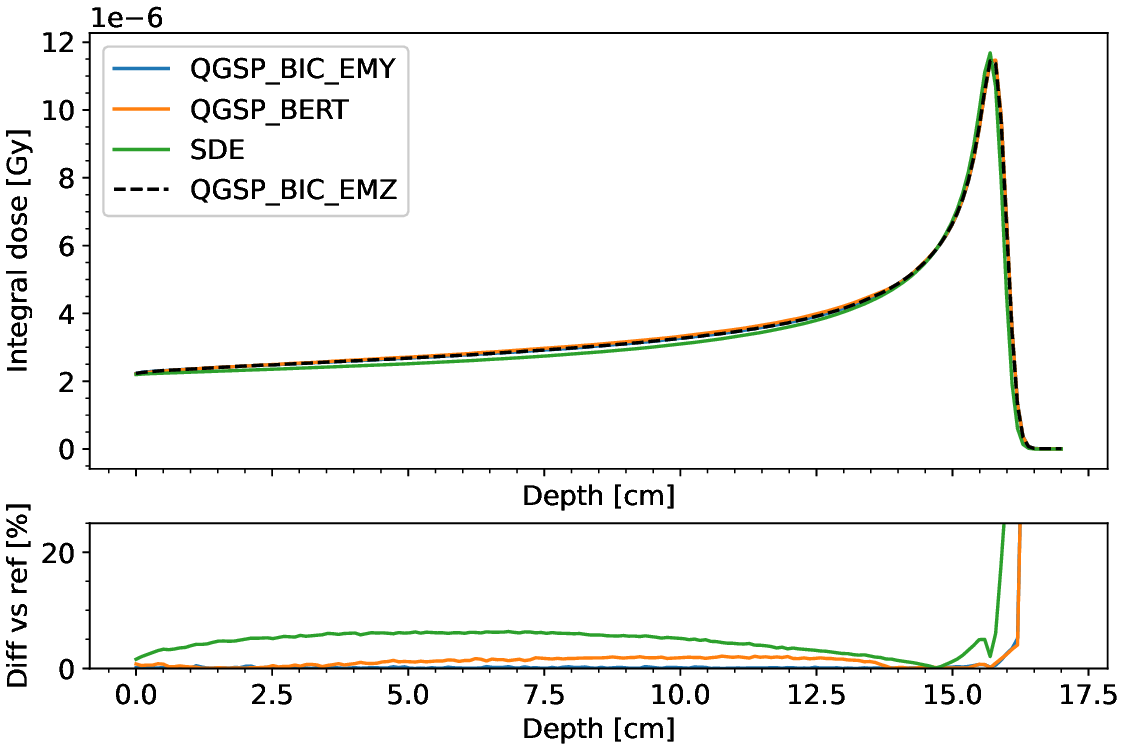}
        \caption{}
        \label{subfig:IDD4}
    \end{subfigure}
    \caption{Integral depth-dose comparison between different Geant4 physical models and the SDE model for two beam energies (a) 100 MeV and (b) 150 MeV. The reference physics list (\texttt{QGSP\_BIC\_EMZ}) is shown in dotted lines.}
    \label{fig:physlistcomp}
\end{figure}

To contextualise the observed differences between the SDE model and Geant4, Figure \ref{fig:physlistcomp} compares integral depth-dose curves obtained with different Geant4 physics lists that are commonly used for proton therapy calculations (\texttt{QGSP\_BIC\_EMZ, QGSP\_BIC\_EMY} and \texttt{QGSP\_BERT}). Comparisons are shown for 100 MeV (Fig.\ \ref{subfig:IDD3}) and 150 MeV (Fig.\ \ref{subfig:IDD4}) proton beams, with the percentage differences calculated with respect to the reference physics list, \texttt{QGSP\_BIC\_EMZ}. The SDE curve is also included for reference, illustrating that the deviation between the SDE model and the reference Geant4 configuration lies within the typical range of variability observed among Geant4 physics lists. 

Notably, while the SDE model exhibits the largest discrepancies in the low-dose tail of the Bragg curve, agreement in the high-dose region near the peak remains within the inter-list variations of Geant4. These results reinforce the physical consistency of the SDE model, while future efforts will focus on reducing these discrepancies through the implementation of secondary particle modelling. Since \texttt{QGSP\_BIC\_EMY} and \texttt{QGSP\_BERT} are computationally less expensive than \texttt{QGSP\_BIC\_EMZ}, it is most meaningful to compare their runtimes with the SDE model. For this geometry, the SDE implementation is from 2.7 to 3.1 times faster than \texttt{QGSP\_BIC\_EMY} and from 2.3 to 2.5 times faster than \texttt{QGSP\_BERT}.

\subsection{Longitudinally heterogeneous slab phantom} \label{sec:longhetero_results}

\begin{figure}[ht!]
    \centering
    \begin{subfigure}[t]{0.48\textwidth}
        \centering
        \includegraphics[width=\textwidth]{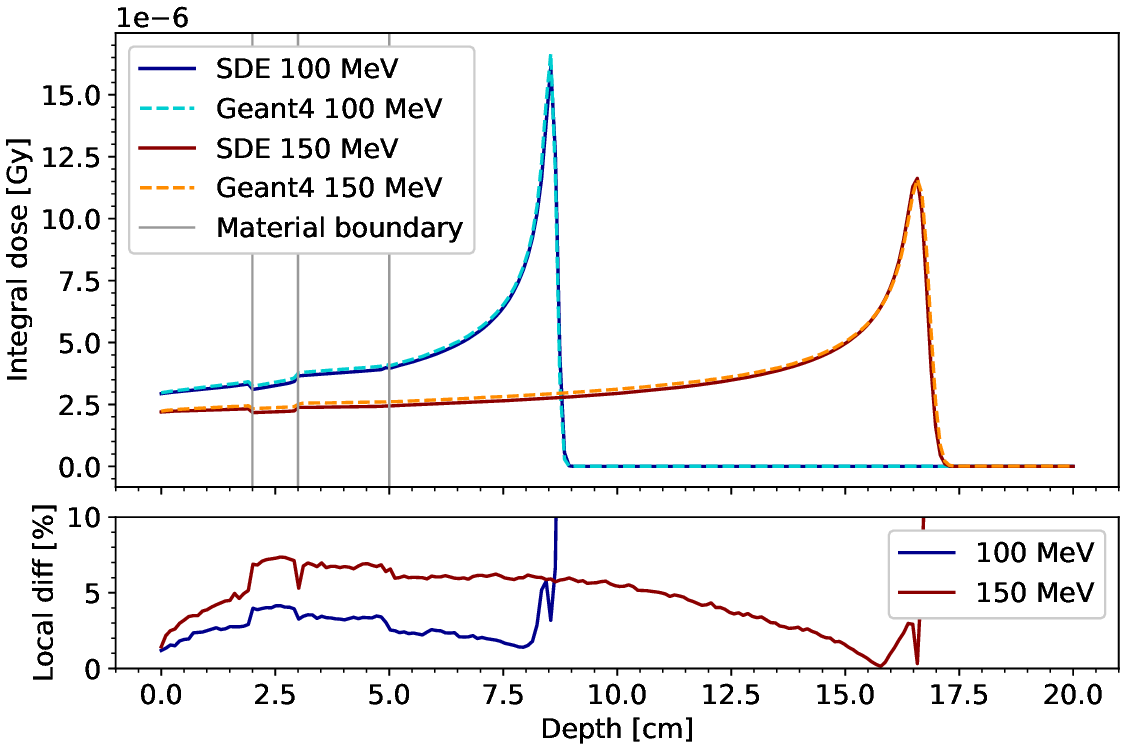}
        \caption{}
        \label{subfig:IDD2}
    \end{subfigure}%
    ~ 
    \begin{subfigure}[t]{0.48\textwidth}
        \centering
        \includegraphics[width=\textwidth]{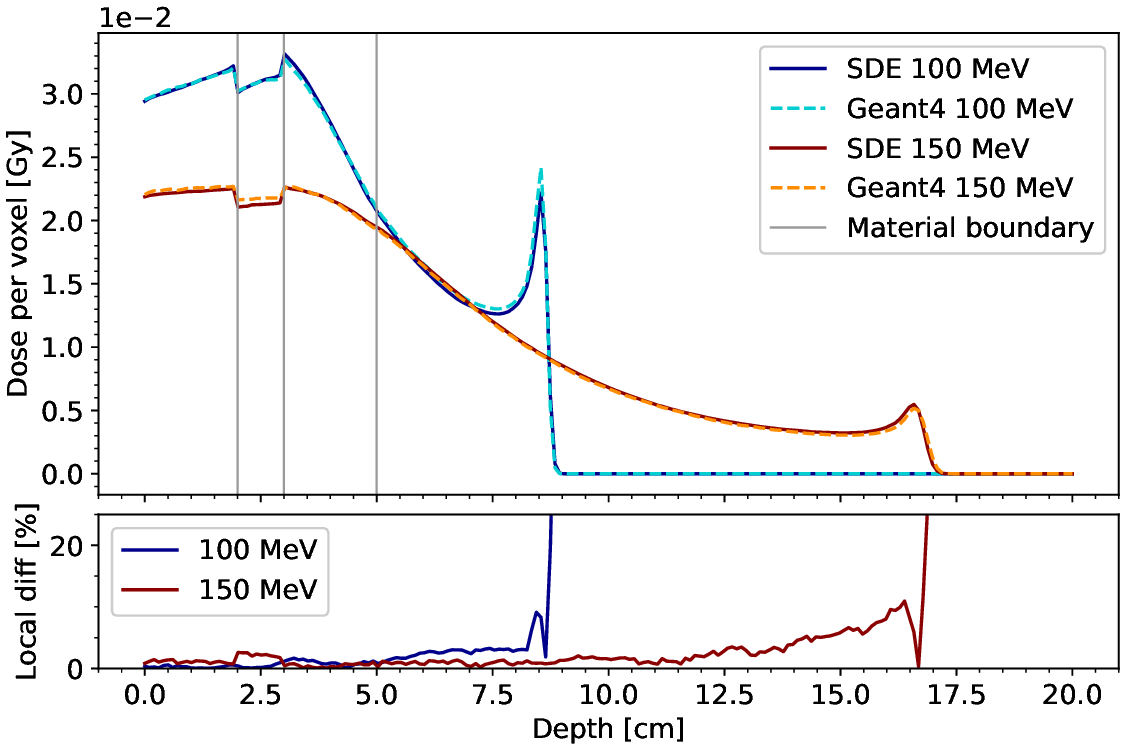}
        \caption{}
        \label{subfig:1Dslice2}
    \end{subfigure}
    \caption{1D comparison between SDE and Geant4 for two monoenergetic proton beams (100 and 150 MeV) in a water phantom with a 1-cm bone layer and a 2-cm low density lung layer, with material boundaries delimited with grey solid lines. (a) Integral depth-dose curves, (b) Central-axis depth-dose curves. Lower subplots show pointwise percentage dose differences.}
    \label{fig:1Dcomp2}
\end{figure}

To assess the capability of the proposed model to handle multiple material interfaces and associated range shifts, a second test was conducted adding a bone slab and a low-density compressed lung slab to the water phantom. All other parameters were kept identical to those of the homogeneous water case described in the previous section. Figure \ref{fig:1Dcomp2} presents the 1D comparisons for 100 and 150 MeV monoenergetic proton beams. The integrated depth–dose curves show excellent agreement between the two models, with calculated range differences below 0.2 mm for 100 MeV and 0.4 mm for 150 MeV. 

\begin{figure}[ht!]
    \centering
    \begin{subfigure}[t]{0.48\textwidth}
        \centering
        \includegraphics[width=\textwidth]{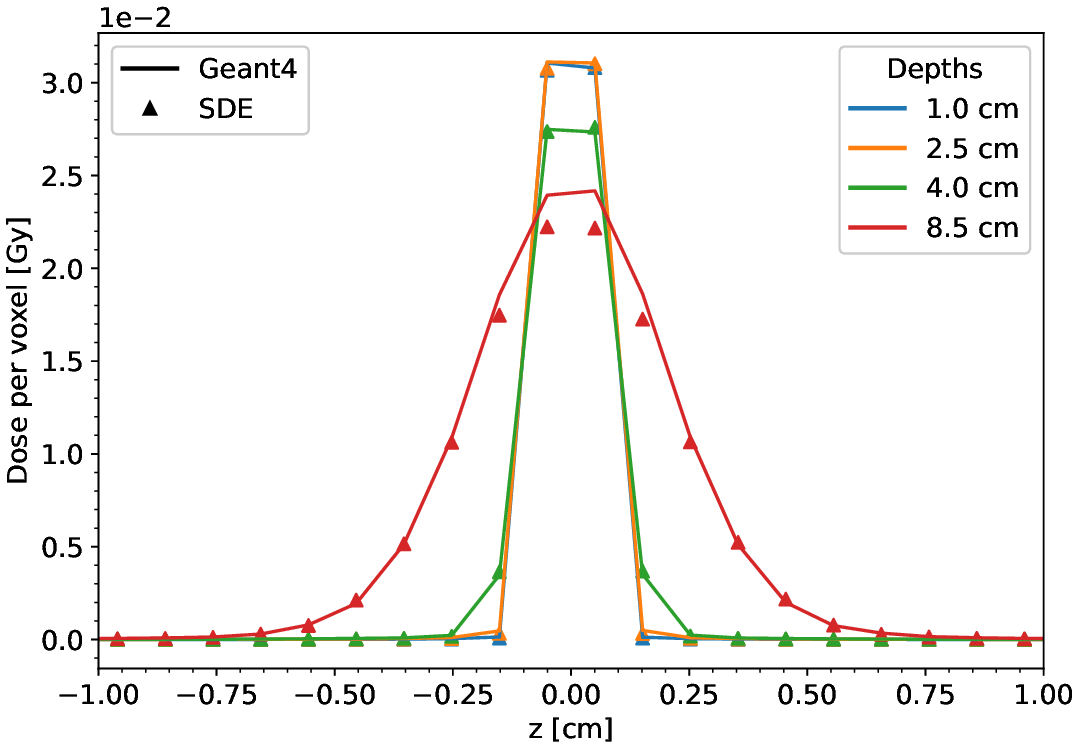}
        \caption{}
        \label{subfig:lateral100MeV_2}
    \end{subfigure}%
    ~ 
    \begin{subfigure}[t]{0.48\textwidth}
        \centering
        \includegraphics[width=\textwidth]{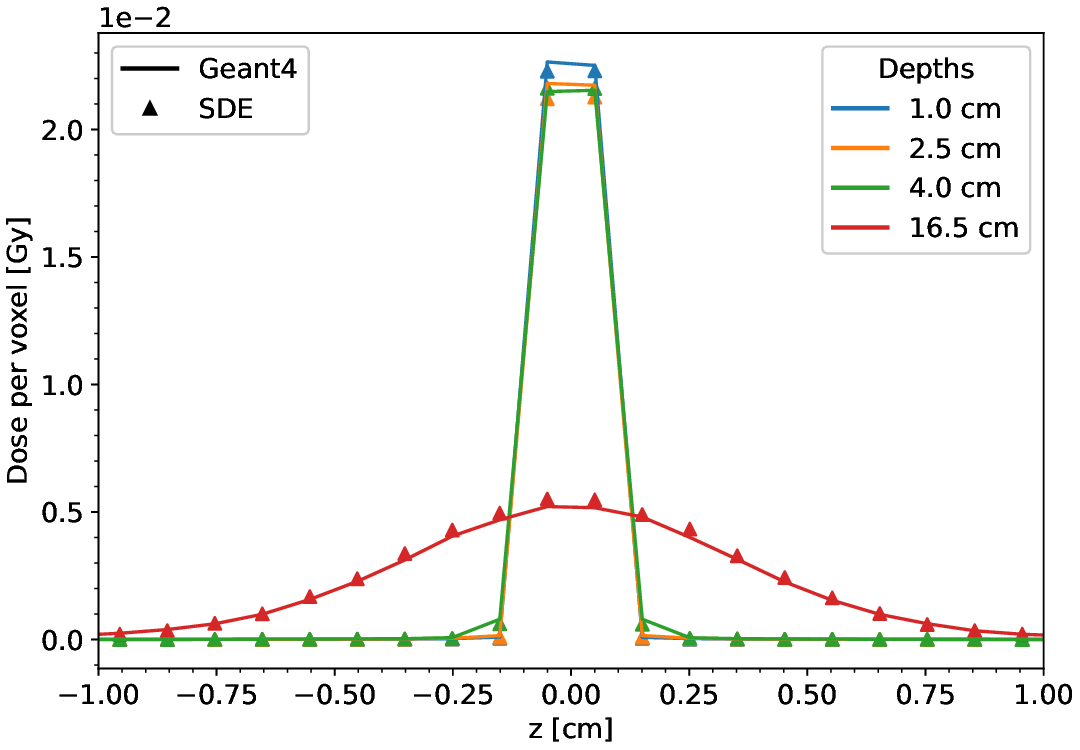}
        \caption{}
        \label{subfig:lateral150MeV_2}
    \end{subfigure}
    \caption{Lateral profile comparison between SDE (marker points) and Geant4 (solid lines) for two proton energies in a longitudinally heterogeneous phantom (a) Using a 100 MeV beam, (b) using a 150 MeV beam. The profiles at 2.5 and 4 cm correspond to the bone and lung regions.}
    \label{fig:1Dprofiles2}
\end{figure}

\begin{figure}[ht!]
    \begin{subfigure}[t]{0.56\textwidth}
        \centering
        \includegraphics[width=\textwidth]{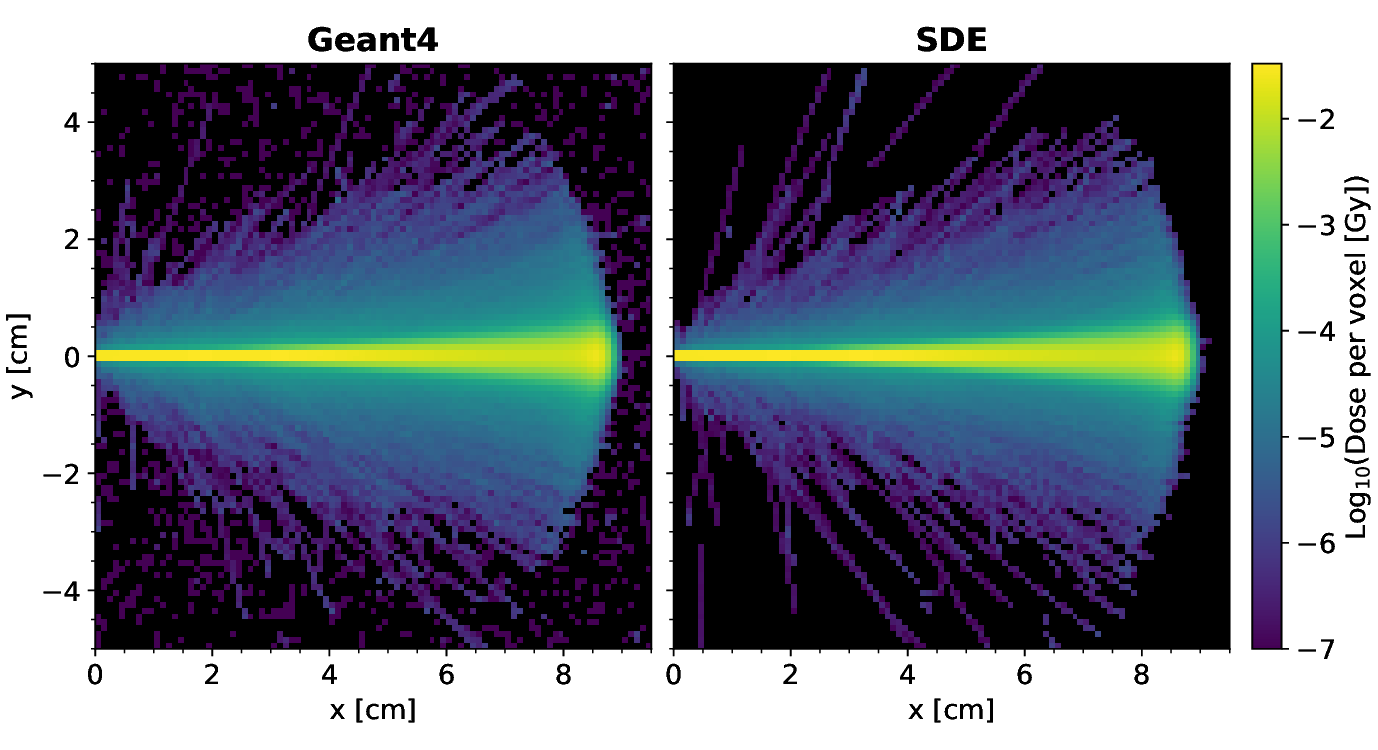}
        \caption{}
        \label{subfig:2Dslice100MeV_slab}
    \end{subfigure}
    \begin{subfigure}[t]{0.43\textwidth}
        \centering
        \includegraphics[width=\textwidth]{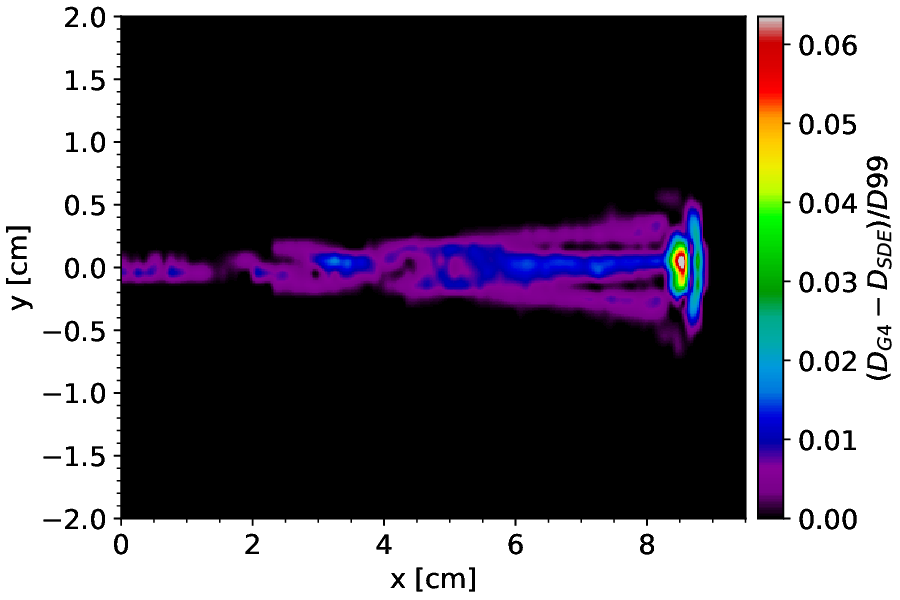}
        \caption{}
        \label{subfig:doseDiff100MeV_slab}
    \end{subfigure}
    \vspace{-2mm}
    \begin{subfigure}[t]{0.56\textwidth}
        \centering
        \includegraphics[width=\textwidth]{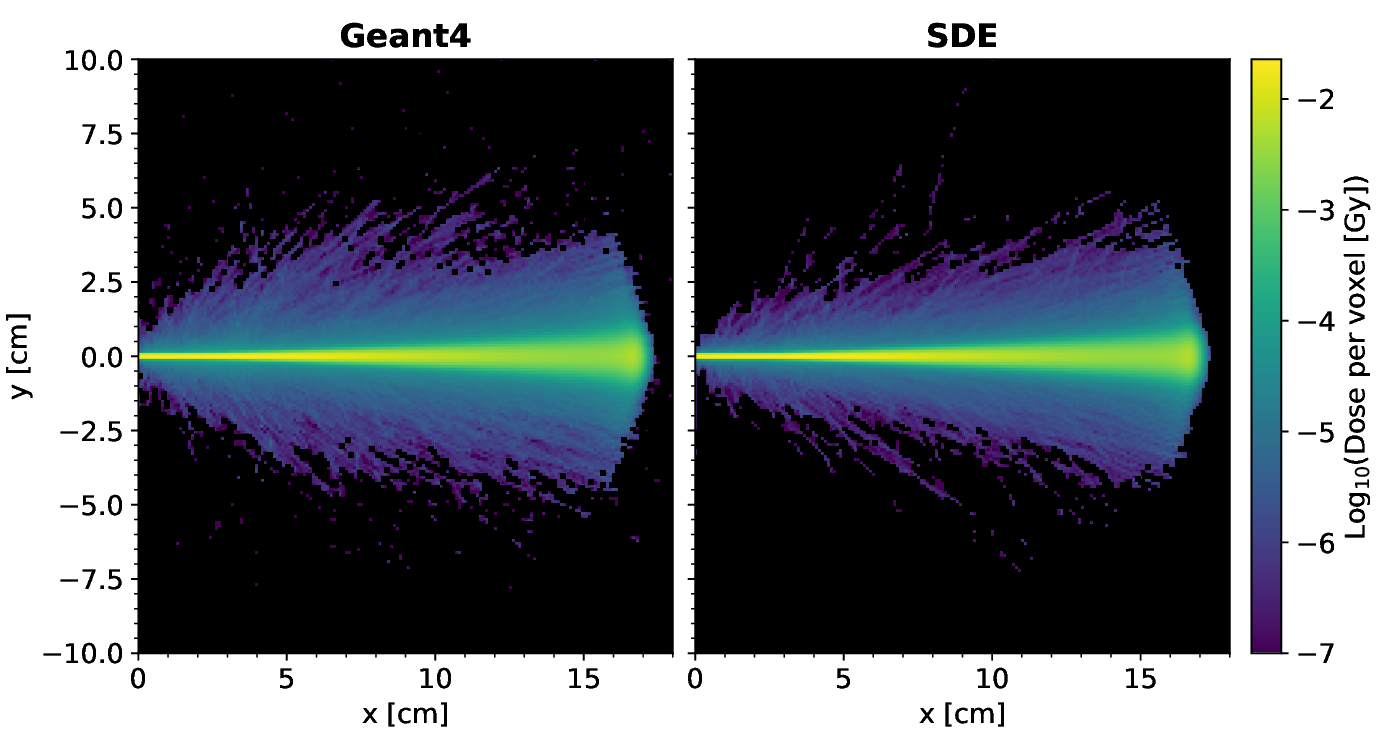}
        \caption{}
        \label{subfig:2Dslice150MeV_slab}
    \end{subfigure}
    \hspace{2mm}
    \begin{subfigure}[t]{0.43\textwidth}
        \centering
        \includegraphics[width=\textwidth]{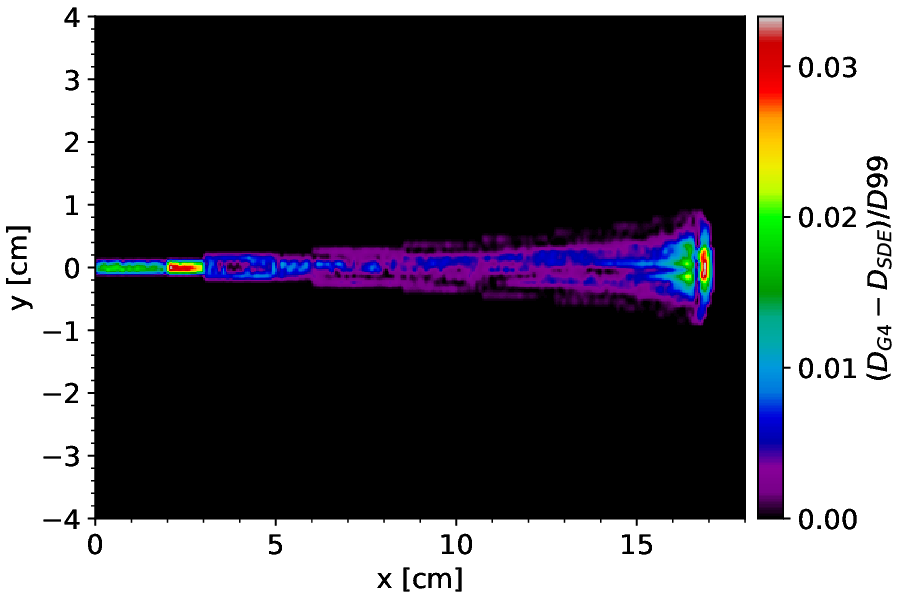}
        \caption{}
        \label{subfig:doseDiff150MeV_slab}
    \end{subfigure}
    \caption{2D dose distribution comparison between the SDE model and Geant4 in a slab phantom for monoenergetic proton beams. (a) Central-axis dose distribution and (b) corresponding relative dose difference map for a 100 MeV beam, normalised to the 99th percentile (D99) of the Geant4 reference dose. (c,d) Same quantities for a 150 MeV beam.}
    \label{fig:2DsliceComp2}
\end{figure}

The lateral dose profiles for both proton energies in the slab phantom are shown in Figure \ref{fig:1Dprofiles2}, whereas the 2D dose distributions and voxel-wise relative difference maps for this scenario are shown in Figure \ref{fig:2DsliceComp2}. Overall, the behaviour is consistent with that observed in the homogeneous phantom, with the SDE model reproducing the lateral profile shapes and widths across all material layers. This indicates that the dominant beam broadening is well described throughout the heterogeneous geometry. Profiles are shown at representative depths within each slab layer, as well as at the depths corresponding to the Bragg peak positions. For the 100 MeV beam, the largest discrepancies are concentrated around the Bragg peak, as also observed in the homogeneous case. In the slab configuration, this difference is more noticeable due to a sharper Bragg peak at lower energy and the presence of multiple material interfaces, which make the peak height more sensitive to small differences in scattering and energy-deposition modelling, even when the proton range is well matched.

For the lower-energy beam, milder differences are observed in the low-density lung region between depths of 3 and 5 cm. In low-density media, energy deposition is more strongly influenced by how energy is redistributed away from the proton track. Geant4 explicitly transports secondary electrons, leading to a slightly broader lateral redistribution of dose, while the SDE model deposits energy more locally. This results in small but spatially coherent differences in the lung region.

For the higher-energy beam, differences in the lung region are less apparent, while the main discrepancies occur within the bone layer between depths of 2 and 3 cm. At higher energies, transport in low-density media becomes less sensitive to secondary-electron effects, whereas the increased stopping power and sharp density change in bone have a greater influence on the dose distribution. As a result, differences between the two models are more visible in the high-density layer.

To further quantify agreement, gamma analysis was performed for both beam energies in the slab configuration, yielding pass rates of 95.1\% for the 100 MeV beam and 98.6\% for the 150 MeV beam using a 2\%/0.5 mm criterion with a 1\% dose threshold. In terms of computational performance, the SDE model was between 2.6 and 3 times faster than single-threaded Geant4 for this geometry.

\subsection{Laterally heterogeneous composite phantom}

\begin{figure}[ht!]
    \centering
    \begin{subfigure}[t]{0.48\textwidth}
        \centering
        \includegraphics[width=\textwidth]{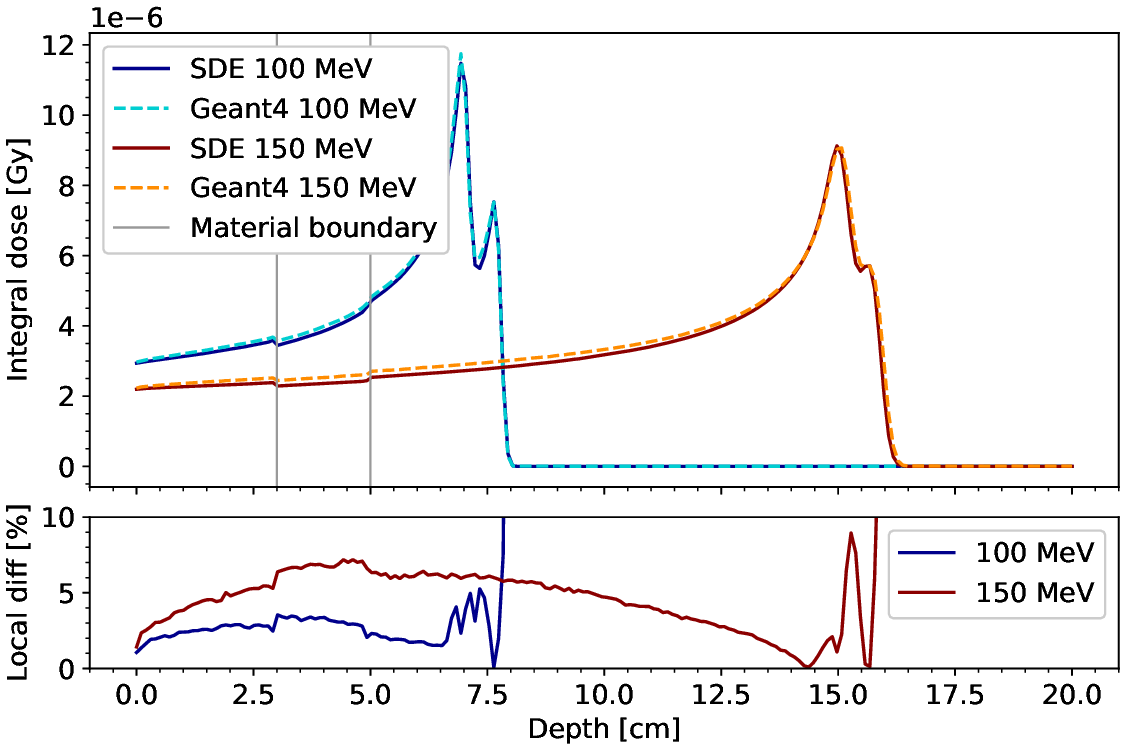}
        \caption{}
        \label{subfig:IDD3_insert}
    \end{subfigure}%
    ~ 
    \begin{subfigure}[t]{0.48\textwidth}
        \centering
        \includegraphics[width=\textwidth]{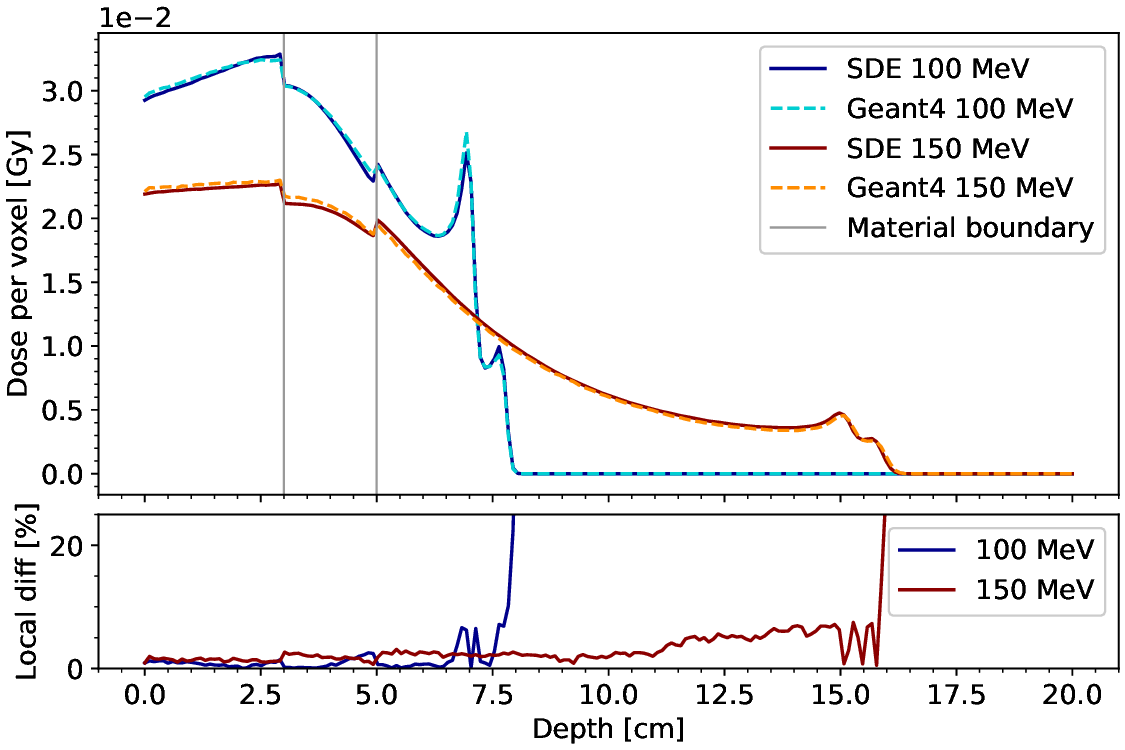}
        \caption{}
        \label{subfig:1Dslice3}
    \end{subfigure}
    \caption{1D comparison between SDE and Geant4 for two monoenergetic proton beams (100 and 150 MeV) in a laterally heterogeneous phantom with a 2-cm off-axis bone insert, with material boundaries delimited with grey solid lines. (a) Integral depth-dose curves, (b) Central-axis depth-dose curves. Lower subplots show pointwise percentage dose differences.}
    \label{fig:1Dcomp3}
\end{figure}

Having established good agreement between the SDE model and Geant4 for the homogeneous and slab phantom configurations, the laterally heterogeneous insert phantom provides a more stringent test of the model’s response to localised lateral heterogeneities, which are known to challenge reduced proton transport models \citep{schaffner1999dose}. Figure \ref{fig:1Dcomp3} shows the integrated and central-axis dose distributions, which remain in close agreement and within the same percentage difference ranges observed for the previous test cases. 
Both models reproduce the two expected high-dose peaks arising from the range shift experienced by the upper half of the beam. The agreement in proton range for the more prominent peak is within 0.2 mm for the 100 MeV case and within 0.5 mm for the 150 MeV case, with a comparable level of agreement retained for the secondary peak.

\begin{figure}[ht!]
    \centering
    \begin{subfigure}[t]{0.48\textwidth}
        \centering
        \includegraphics[width=\textwidth]{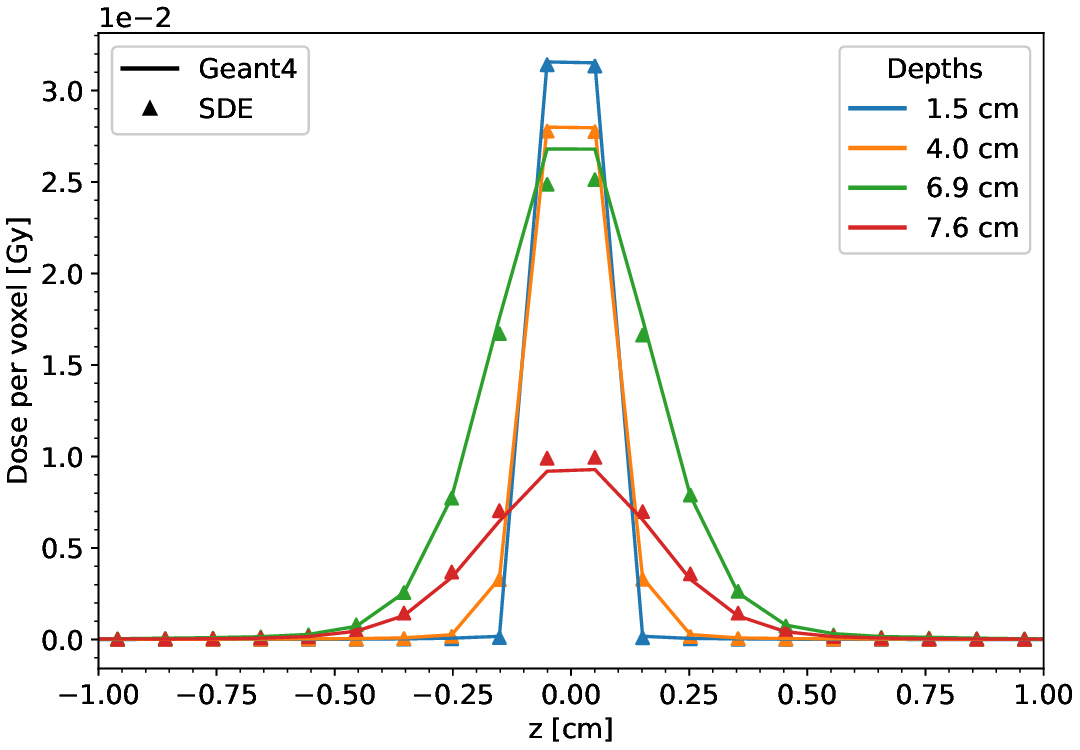}
        \caption{}
        \label{subfig:lateral100MeV_3}
    \end{subfigure}%
    ~ 
    \begin{subfigure}[t]{0.48\textwidth}
        \centering
        \includegraphics[width=\textwidth]{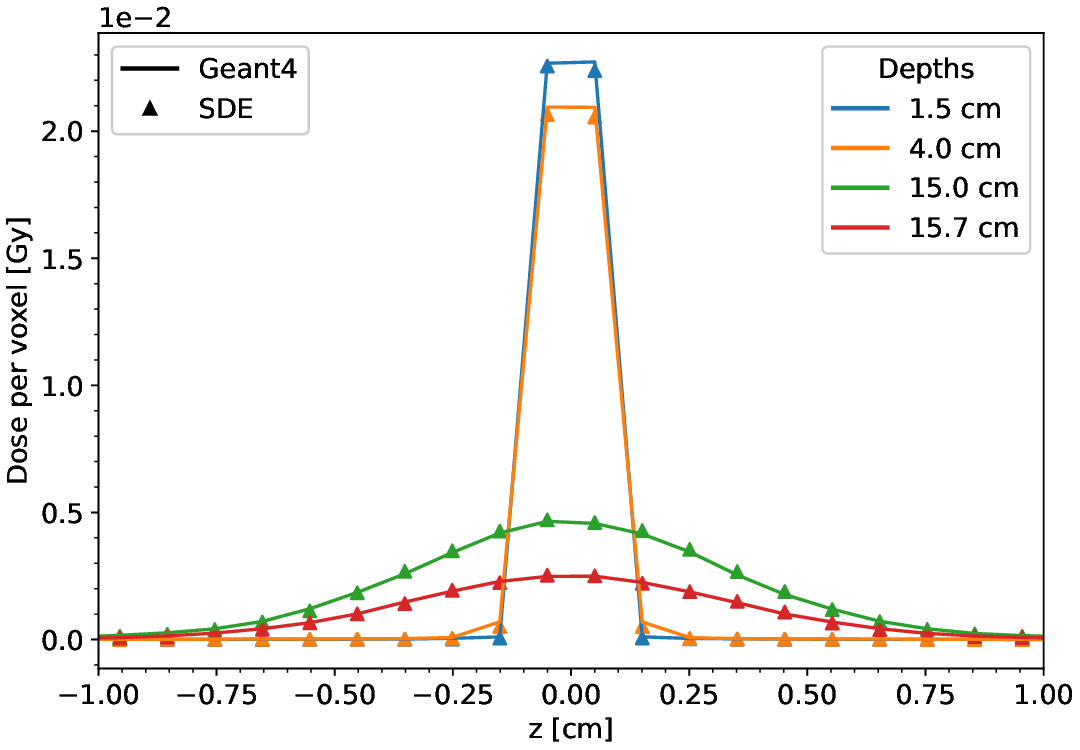}
        \caption{}
        \label{subfig:lateral150MeV_3}
    \end{subfigure}
    \caption{Lateral dose profile comparison between the SDE model (marker points) and Geant4 (solid lines) for two proton energies (a) 100 MeV and (b) 150 MeV in a laterally heterogeneous phantom at different depths. The profile at 4 cm corresponds to the bone insert region. The profiles at the two largest depths correspond to the locations of the high-dose peaks observed in the 1D depth–dose distributions.}
    \label{fig:1Dprofiles3}
\end{figure}

Lateral dose profiles are shown in Figure \ref{fig:1Dprofiles3} and illustrate the influence of the bone insert at greater depths. The two shallower profiles correspond to the entrance water region and the bone insert location, while the deeper profiles correspond to the two Bragg peak positions formed by the upper and lower portions of the beam. While the overall lateral extent of the dose distributions is comparable between the SDE model and Geant4, small differences are observed in the central region of the profiles. For the 100 MeV beam, a reduced dose is again observed for one of the Bragg peaks, whereas the secondary peak shows closer agreement, consistent with its broader spatial distribution. These differences reflect variations in the radial redistribution of deposited energy in the presence of localised heterogeneities, rather than changes in the overall lateral spread of the beam.

\begin{figure}[h!]
    \begin{subfigure}[t]{0.56\textwidth}
        \centering
        \includegraphics[width=\textwidth]{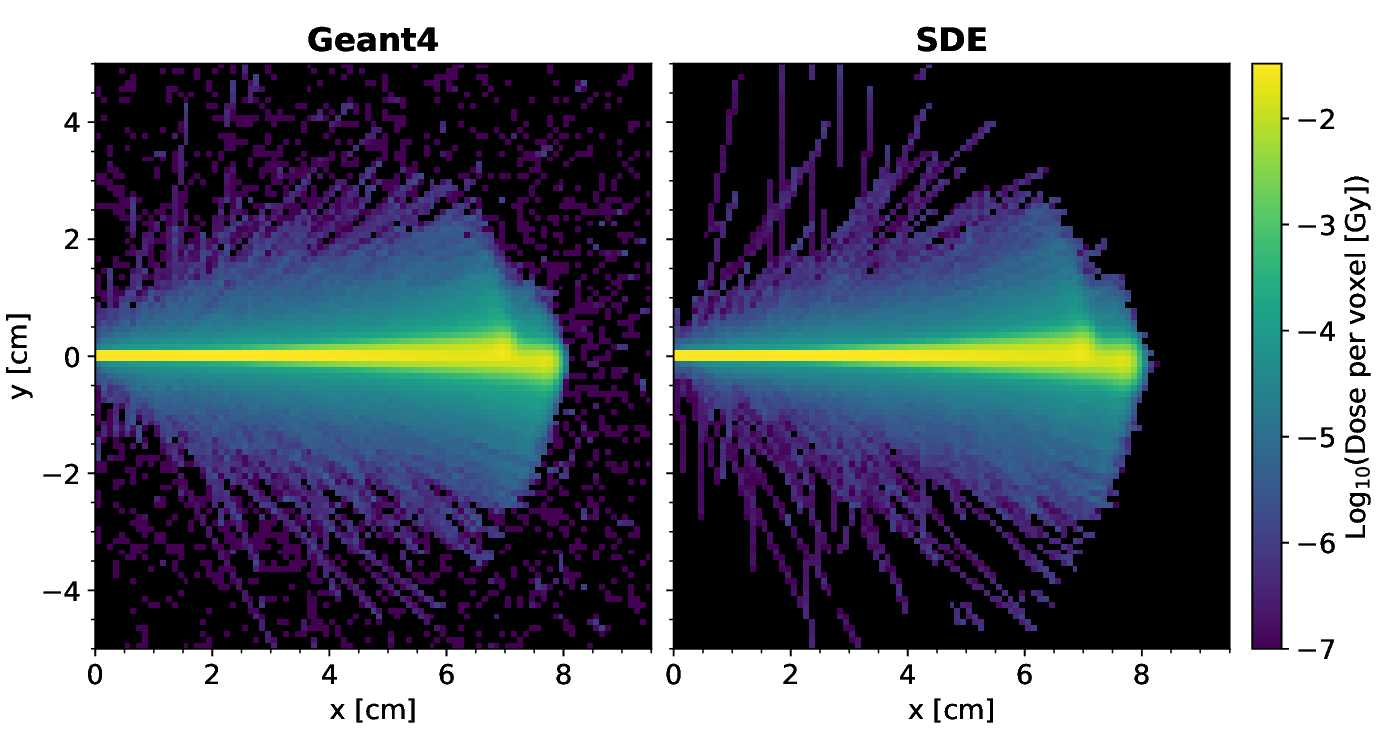}
        \caption{}
        \label{subfig:2Dslice100MeV_insert}
    \end{subfigure}
    \begin{subfigure}[t]{0.43\textwidth}
        \centering
        \includegraphics[width=\textwidth]{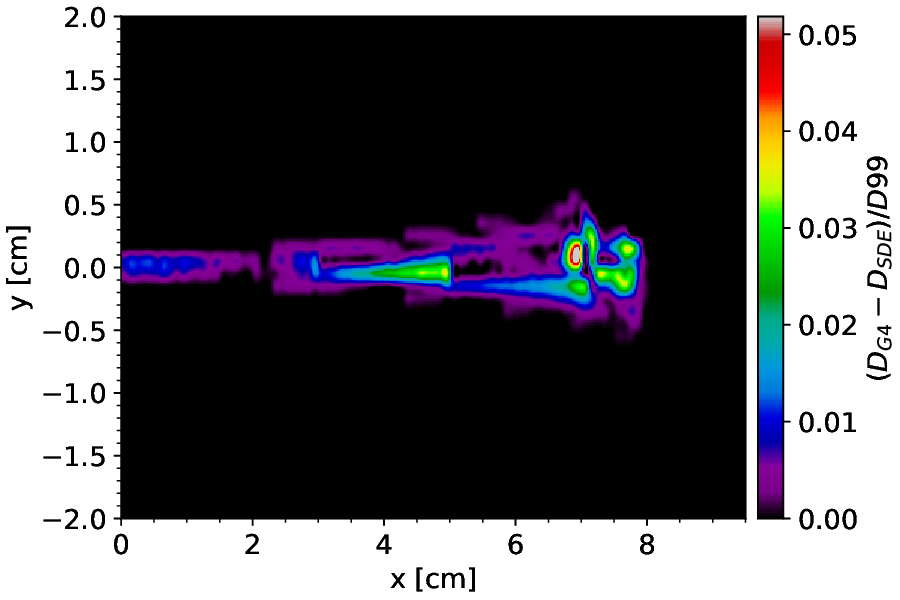}
        \caption{}
        \label{subfig:doseDiff100MeV_insert}
    \end{subfigure}
    \vspace{-2mm}
    \begin{subfigure}[t]{0.56\textwidth}
        \centering
        \includegraphics[width=\textwidth]{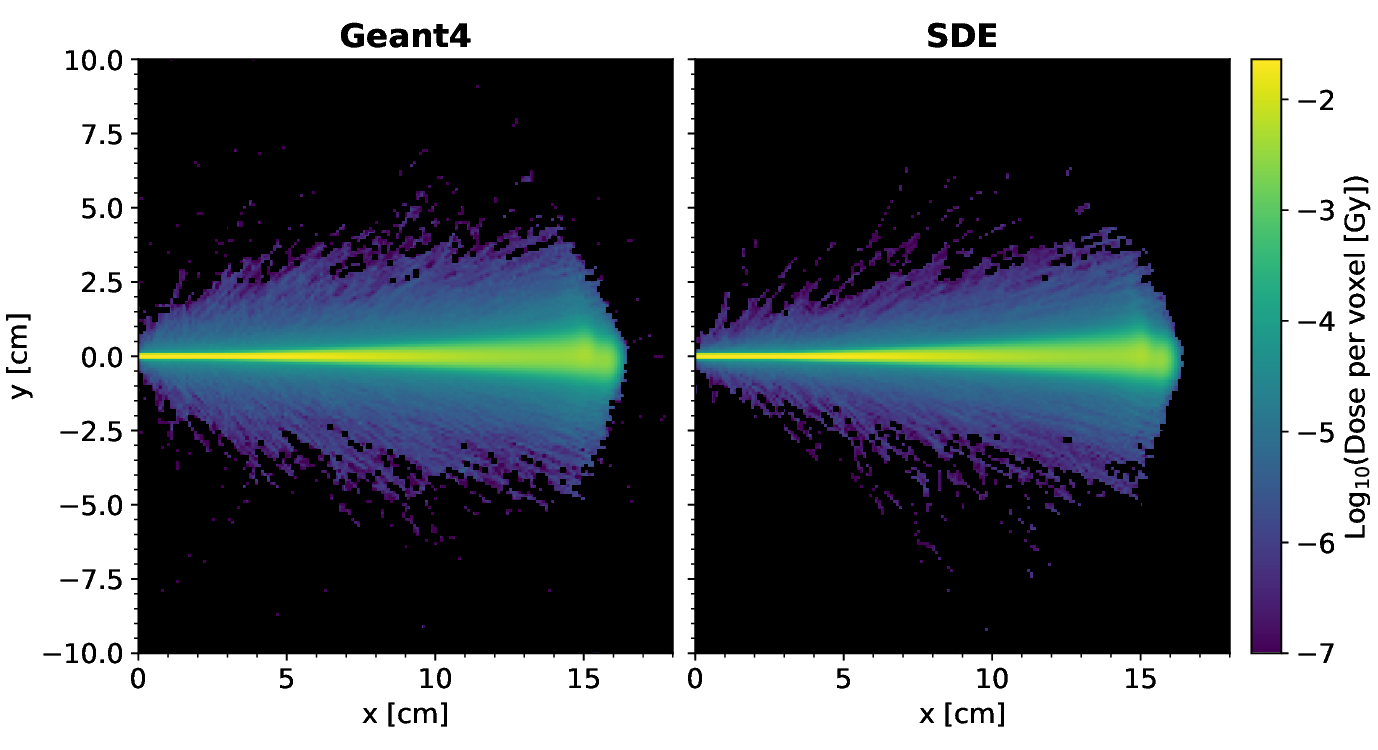}
        \caption{}
        \label{subfig:2Dslice150MeV_insert}
    \end{subfigure}
    \hspace{2mm}
    \begin{subfigure}[t]{0.43\textwidth}
        \centering
        \includegraphics[width=\textwidth]{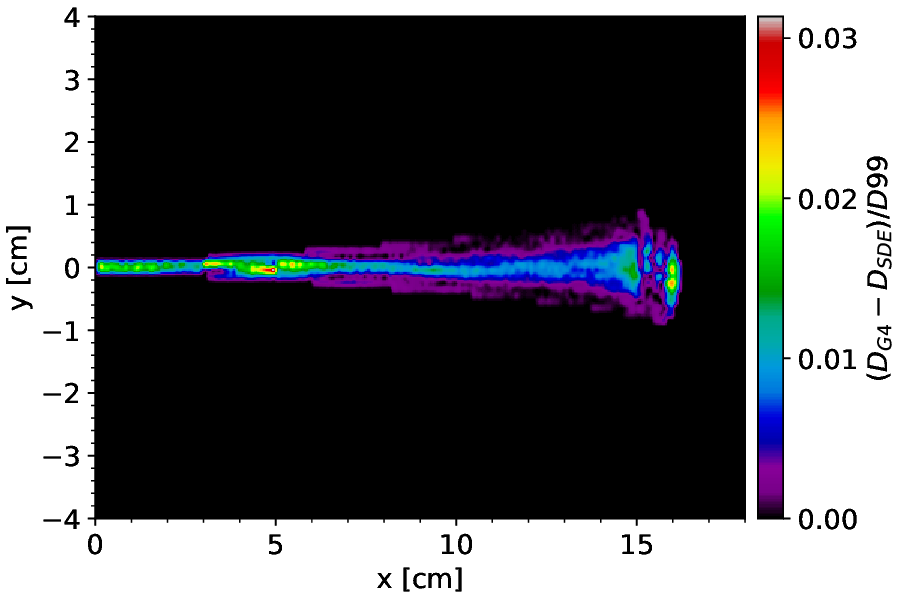}
        \caption{}
        \label{subfig:doseDiff150MeV_insert}
    \end{subfigure}
    \caption{2D dose distribution comparison between the SDE model and Geant4 in a water phantom with an off-axis bone insert for monoenergetic proton beams. (a) Central-axis dose distribution and (b) corresponding relative dose difference map for a 100 MeV beam, normalised to the 99th percentile (D99) of the Geant4 reference dose. (c,d) Same quantities for a 150 MeV beam.}
    \label{fig:2DsliceComp3}
\end{figure}

The 2D dose distributions and voxel-wise relative difference maps for this configuration are shown in Figure \ref{fig:2DsliceComp3}. The effects of lateral heterogeneities are well captured by the SDE model, with relative differences remaining within the same limits observed for the homogeneous and slab phantoms. This indicates that the magnitude and spatial localisation of the model deviations are stable across the different phantom geometries considered.

Gamma analysis for the insert phantom yielded pass rates of 96.2\% for the 100 MeV beam and 98.2\% for the 150 MeV beam, using the same stringent criteria applied in the previous scenarios. In terms of computational performance, the SDE model remains between 2.5 and 2.9 times faster than single-threaded Geant4 for this configuration.

\section{Discussion and Conclusion}

This study presents the first systematic validation of the SDE-based proton transport model against Geant4 across homogeneous and heterogeneous phantom geometries. The results demonstrate that the SDE model reproduces the principal dosimetric features predicted by Geant4 with consistent accuracy. Proton range agreement remained within 0.2 mm for 100 MeV beams and within 0.6 mm for 150 MeV beams across all test cases, and gamma pass rates exceeded 95\% under stringent 2\%/0.5 mm criteria with a 1\% dose threshold---well within clinically acceptable bounds. Importantly, the observed differences between the two models were spatially stable and well characterised: discrepancies were concentrated around steep dose gradients and material interfaces, while entrance, plateau, and lateral regions showed close agreement. Comparison against multiple Geant4 physics lists further confirmed that the deviations introduced by the SDE model lie within the range of inter-list variability typical of Geant4 itself, reinforcing the physical consistency of the approach. The computational performance of the SDE model was also stable across all configurations, achieving speed-ups of approximately 2.5--3× relative to single-threaded Geant4, with the greatest gains observed at higher beam energies. It is important to frame the claim being made here precisely: the SDE formulation offers a genuinely novel application of stochastic differential equations to charged particle transport, and--- independent of any hardware acceleration---it can simulate proton transport faster than conventional Monte Carlo codes for comparable dosimetric accuracy. While GPU-accelerated Monte Carlo engines are increasingly being adopted in clinical practice, a factor of approximately three in per-proton computational cost represents a meaningful qualitative improvement. Such gains could enable longer optimisation runs during treatment planning on more modest computing infrastructure that not all clinics have the ability to realise.

The simplifications adopted in the current model---notably the absence of secondary particle transport and the use of a fixed energy absorption threshold---were shown to be appropriate for producing clinically relevant outputs in the geometries and energy ranges considered. The largest resulting effects were a steeper distal fall-off and the absence of low-level dose contributions at remote locations, both of which are expected and well understood. Regarding model parameterisation, the central-axis depth–dose curves exhibited sensitivity to the angular cutoffs used for Rutherford and backscatter events, which are currently set as fixed values. Investigating energy-dependent cutoffs in future work may reduce the number of free parameters and improve robustness across a wider range of clinical scenarios. Beyond the single-threaded results reported here, the SDE framework is naturally amenable to further acceleration. Each sequential proton track evolves independently, making the model straightforward to parallelise across multiple CPU cores---a strategy routinely used to accelerate Monte Carlo simulations---or to port to GPU architectures. Aside from relatively rare inelastic proton-nucleus collision which can cause macroscopic losses in proton energy, each SDE step involves identical arithmetic operations across all protons, making the method well-suited to SIMD (Single Instruction, Multiple Data) execution. This suggests substantial additional speed improvements are achievable without fundamental algorithmic modifications, and positions the SDE framework as a natural candidate for GPU acceleration.

Additionally, it is worth mentioning that the reported dose corresponds to dose-to-medium, obtained by dividing the deposited energy by the local voxel mass. In heterogeneous materials, dose-to-medium differs from dose-to-water due to material-dependent stopping powers and nuclear interactions, as discussed by \cite{paganetti2009dose}. In the present study, all comparisons between the SDE model and Geant4 are performed consistently using dose-to-medium. For potential use within treatment planning workflows, conversion to dose-to-water could be performed using established stopping-power–based formalisms such as that of Paganetti.

Several avenues for future development follow from the present work. Incorporating simplified models for secondary particle contributions, particularly neutrons and gamma rays, would improve agreement in the distal tail and at remote locations. Extending the validation to patient-derived CT geometries and spread-out Bragg peaks would provide a more complete assessment of clinical applicability. Such geometries will introduce additional challenges, including voxelised material assignment and the presence of air cavities with sharp density contrasts. The phantom-based validation presented here establishes the necessary foundation by demonstrating accuracy across controlled homogeneous and heterogeneous configurations before confronting these complexities. Beyond computational speed, the continuous and differentiable structure of the SDE formulation opens possibilities that are fundamentally difficult to achieve with discrete Monte Carlo sampling. The well-characterised probabilistic structure of the SDE provides a natural framework for propagating input uncertainties (such as stopping power ratios or patient positioning errors) directly through the transport model, complementing conventional scenario-based robustness evaluation strategies, e.g. optimising the worst-case scenario. Realising these capabilities in practice will require further methodology, but the mathematical foundation provided by the SDE framework makes such extensions feasible. Source code optimisation and integration with treatment planning workflows are also natural next steps. Overall, these findings establish the SDE approach as a promising, fast, and accurate complement to conventional Monte Carlo methods for proton therapy dose calculation.

\appendix
\section{Appendix}
\subsection{Angle conversion between reference frames}
\label{appendix:angle_conv}
During the collision of a two-body system between a proton of mass $m_1$ and a target nucleus of mass $m_2$, the angular distribution of the outgoing proton is obtained from nuclear data, which is often reported in the center of mass (CM) reference frame. Hence, it is necessary to perform a transformation of the angles to the lab (L) frame for use in the present model. Let $v_{1L}$ and $v_{1C}$ be the initial proton velocities in the lab and CM frames respectively, and $v_{1L}'$ and $v_{1C}'$ the final velocities in the respective frames. In general, the expression for the angle $\theta_L$ in the lab frame in terms of the angle $\theta_{CM}$ in the center of mass frame, bearing in mind relativistic effects, is given by

\begin{equation}
\label{eq:cm_to_lab}
    \tan{\theta_L} = \frac{\sin{\theta_{CM}}}{ \gamma_u \left( \cos{\theta_{CM}} + \frac{u}{v_{1C}'} \right)},
\end{equation}
where $u$ is the velocity of the center of mass of the two-body system and $\gamma_u = 1/\sqrt{1-u^2}$. If the  kinetic energy $K$ of the proton is known, then its total energy $E_{1L}$ and momentum $p_{1L}$ can be calculated using
\begin{align*}
    E_{1L} &= K + m_1, \\
    p_{1L} &= \sqrt{E_{1L}^2 - m_1^2}.
\end{align*}
In addition, $u$ is also defined from these quantities, as it is not dependent on the nature of the collision:
\begin{equation*}
u = \frac{p_{tot}}{E_{tot}}=\frac{p_{1L}}{E_{1L}+ m_2}.
\end{equation*}

Given that we need an expression for the final proton velocity in the CM frame, we use the Lorentz velocity transformation
\begin{equation*}
    v_{1C}' = \frac{v_{1L}'-u}{1-uv_{1L}'}= \frac{p_{1L}'-uE_{1L}'}{E_{1L}'-up_{1L}'}.
\end{equation*}
For the elastic case, the proton velocity remains the same before and after the collision in the CM frame, thus, we may use $p_{1L}'=p_{1L}$ and $E_{1L}'=E_{1L}$ with the definitions given earlier. For the inelastic case, there will be an associated proton energy loss $E^*$, meaning that
\begin{align*}
    E_{1L}'&=E_{1L}-E^*, \\
    p_{1L}'&=\sqrt{E_{1L}'^2-m_1^2}.
\end{align*}
The outgoing proton energy can also be obtained from the nuclear database in the CM frame (referred to as $E_{1C}'$). To transform the energy back to the lab frame, the inverse Lorentz transformation can be used:
\begin{equation}
    E_{1L}' = \gamma_u \left(E_{1C}'+u\cos(\theta_{CM})\sqrt{(E_{1C}')^2-m_1^2} \right).
\end{equation}

\subsection{Integration of \eqref{eq:large_elastic_cdf} for hydrogen elastic scattering}
Firstly, to integrate \(\sigma_{R}^c(\mu, E)\) we use that the anti-derivative of \(\sigma_{R}^c(\mu, E)\), denoted \(\sigma_{R}^{c,(1)}(\mu, E)\), is given by
\begin{equation*}
    \sigma_{R}^{c,(1)}(\mu, E) = \frac{2\eta^2 \mu}{k^2(1-\mu^2)}-\frac{\eta}{2k^2}\sin\left(2\eta\ln\frac{1+\mu}{1-\mu}\right).
\end{equation*} 
Next we move onto integrating \( \sigma_{N}^c(\mu,E)\). The first term of \eqref{eq:hydrogen_not_ruther} is the sum of polynomial functions whose integral is clear. For the second term, first note that by repeated integration by parts, for any \(n+1\)-fold integrable function \(f\), 
\begin{equation*}
    \int_0^\nu f(\mu)\mu^n\mathrm{d}\mu=(-1)^{n+1}n!f^{(n+1)}(0)+\sum_{i=0}^n\frac{(-1)^if^{(i+1)}(\nu)\nu^{n-i}n!}{(n-i)!},
\end{equation*}
where \(f^{(i)}\) is the \(i\)th anti-derivative of \(f\). This and the fact that \(\sigma_N^c\) is even in \(\mu\) implies the integral of the second term can be written in terms of the antiderivatives of
\begin{equation*}
    f^{\cos}_{\pm}(\mu) =\frac{1}{1\pm\mu}\cos\left(\eta\ln\frac{1\pm\mu}{2}+C\right),
\end{equation*}
where the constant \(C\) is precisely the polar angle of the coefficient \(a_l(E)\). More generally, let
\begin{equation*}
   f_{\pm}(\mu) =\frac{c_1}{1\pm\mu}\cos\left(\eta\ln\frac{1\pm\mu}{2}+C\right)+\frac{c_2}{1\pm\mu}\sin\left(\eta\ln\frac{1\pm\mu}{2}+C\right).
\end{equation*} Using a \(u=\ln((1\pm\mu)/2)\) substitution along with integration by parts we obtain the recursive formula
\begin{align*}
    &f^{(n)}_{\pm}(\mu) ={(1\pm\mu)^{n-1}}\left(c_{\pm,n}^{\cos}\cos\left(\eta\ln\frac{1\pm\mu}{2}+C\right)+c_{\pm,n}^{\sin}\sin\left(\eta\ln\frac{1\pm\mu}{2}+C\right)\right),\\
    &c_{\pm,n+1}^{\cos}=\frac{\pm(nc_{\pm,n}^{\cos}-\eta c_{\pm,n}^{\sin})}{(n^2+\eta^2)},\quad c_{\pm,n+1}^{\sin}=\frac{\pm(\eta c_{\pm,n}^{\cos}+n c_{\pm,n}^{\sin})}{(n^2+\eta^2)},\\
    &c_{\pm,0}^{\cos}=c_1,\quad c_{\pm,0}^{\sin}=c_2.
\end{align*}

\section*{Acknowledgements and data sharing} The authors acknowledge support from the EPSRC grant {\it Mathematical Theory of Radiation Transport: Nuclear Technology Frontiers (MaThRad)}, EP/W026899/2. Moreover, we would like to thank colleagues from the UCLH proton beam facility (particularly Colin Baker and Sarah Osman), from the University of Cambridge (particularly Theophile Bonnet, Valeria Raffuzzi and Esmae Woods), as well Ana Louren\c{c}o from the National Physical Laboratory for help and guidance. 

Data sharing is not applicable to this article as no new data were generated or analysed.

\section*{Supplementary material}
All of the results presented in this work can be replicated using the code available at \url{https://github.com/JereKoskela/proton-beam-sde}, which contains the SDE model source, the Geant4 models and Python scripts to generate the figures.

\bibliography{references}{}
\bibliographystyle{agsm}

\end{document}